\documentclass[journal]{IEEEtran}

\usepackage[noadjust]{cite}
\usepackage{pslatex} %
\usepackage{booktabs}

\usepackage{graphicx}
\usepackage{xcolor}
\usepackage{multirow}

\usepackage{amsmath}
\usepackage{amssymb}
\usepackage{amsfonts} %
\usepackage{calligra} %
\usepackage{mathtools}

\usepackage{amsthm} %

\usepackage{dblfloatfix} %

\usepackage{tikz}
\usepackage{verbatim}
\usetikzlibrary{shapes,backgrounds}
\usetikzlibrary{arrows,fit,automata,positioning,decorations,calc}
\usetikzlibrary{spy}
\usetikzlibrary{matrix,chains,decorations.pathreplacing}

\makeatletter
\let\MYcaption\@makecaption
\makeatother

\usepackage[]{subcaption}

\makeatletter
\let\@makecaption\MYcaption
\makeatother

\usepackage{comment}
\usepackage{listings}

\usepackage{pgfplots}
\usepackage{pgfplotstable}

\usetikzlibrary{patterns}

\usepackage[percent]{overpic}

\usepackage[english]{babel}
\usepackage{blindtext}

\usepackage{acronym}

\newcommand\defeq{\stackrel{\mathclap{\scriptsize\mbox{def}}}{=}}

\usepackage{listings}
\usepackage{float}
\usepackage{textcomp}

\newcommand{\todo}[1]{{\color{orange}(TODO: #1)}}

\newcommand{\srinivas}[1]{{\color{blue} Srinivas: #1}}

\theoremstyle{definition}
\newtheorem{definition}{Definition}[section] %
\theoremstyle{example}
\newtheorem{example}[definition]{Example} %
\theoremstyle{theorem}

\newtheorem{remark}[definition]{Remark}

\usepackage{acronym}
\usepackage{algorithm}
\usepackage{algorithmicx}
\usepackage{algpseudocode}
\usepackage{adjustbox}
\definecolor{mygreen}{rgb}{0,0.6,0}
\definecolor{mygray}{rgb}{0.5,0.5,0.5}
\definecolor{mymauve}{rgb}{0.58,0,0.82}

\acrodef{WCRT}{Worst Case Reaction Time}
\acrodef{WCET}{Worst Case Execution Time}
\acrodef{SOT}{Start Of Tick}
\acrodef{EOT}{End Of Tick}
\acrodef{CPS}{Cyber-Physical Systems}
\acrodefplural{CPS}[CPSs]{Cyber-Physical Systems}
\acrodef{AI}{Artificial Intelligence}

\acrodef{ESS}{Energy Storage System}
\acrodef{SoC}{State of Charge}
\acrodef{MBD}{Model-Based Design}
\acrodef{FPGA}{Field-Programmable Gate Array}

\acrodef{CFG}{Control Flow Graph}
\acrodef{TCFG}{Timed Control Flow Graph}
\acrodef{TCCFG}{Timed Concurrent Control Flow Graph}
\acrodef{TCA}{Tick Cost Automata}

\acrodef{SA}{Security Automata}
\acrodef{SMT}{Satisfiability Modulo Theory}

\acrodef{EV}{Electric Vehicle}
\acrodef{ESS}{Energy Storage System}

\acrodef{ReLU}{Recitifed Linear Units}

\acrodef{ML}{Machine Learning}
\acrodef{NN}{Neural Network}
\acrodef{DNN}{Deep Neural Network}
\acrodef{SNN}{Synchronous Neural Network}
\acrodef{CNN}{Convolutional Neural Network}
\acrodef{SANN}{Synchronous Artificial Neural Network}
\acrodef{SCNN}{Synchronous Convolutional Neural Network}
\acrodef{SNN}{Synchronous Neural Network}
\acrodef{SSNN}{Safe Synchronous Neural Network}
\acrodef{ENN}{Enforced Neural Network}
\acrodef{ANN}{Artificial Neural Network}
\acrodef{RNN}{Recurrent Neural Network}
\acrodef{MLP}{Multi-layer Perceptron}
\acrodef{MNN}{Meta Neural Network}
\acrodef{MNN2C}{Meta Neural Network to C}

\acrodef{AV}{Autonomous Vehicle}
\acrodef{LiDAR}{Light Detection and Ranging}

\acrodef{VOC}{Visual Object Classes}
\acrodef{GTSRB}{German Traffic Sign Recognition Benchmark}

\acrodef{VDTA}{Valued Discrete Timed Automata}
\acrodef{DTA}{Discrete Timed Automata}
\acrodef{RE}{Runtime Enforcement}
\acrodef{RA}{Runtime Assurance}
\acrodef{RV}{Runtime Verification}
\acrodef{RI}{Runtime Interchange}
\acrodef{SA}{Security Automata}
\acrodef{OS}{Operating System}

\acrodef{OH}{Overhead}
\newcommand{\red}[1]{\textcolor{red}{#1}}

\DeclareSymbolFont{bbold}{U}{bbold}{m}{n}
\DeclareSymbolFontAlphabet{\mathbbold}{bbold}

\newcommand{\ignore}[1]{{}}

\newcommand{\bbb}{\mathbb{B}}
\newcommand{\bbn}{\mathbb{N}}

\newcommand{\calL}{{\mathcal L}}
\newcommand{\calP}{{\mathcal P}}
\newcommand{\calA}{{\mathcal A}}
\newcommand{\calD}{{\mathcal D}}

\newcommand{\true}{\ensuremath{\mathsf{true}}}
\newcommand{\false}{\ensuremath{\mathsf{false}}}
\newcommand{\pref}{\preccurlyeq}
\newcommand{\sem}[1]{[\!\left[#1\right]\!]}

\newcommand{\ef}{\ensuremath{E_{\varphi}}}

\newcommand{\editI}{\mathsf{editI_{\varphi_I}}}
\newcommand{\editO}{\mathsf{editO_{\varphi}}}
\newcommand{\editIaut}{\mathsf{editI_{\calA_{\varphi_I}}}}
\newcommand{\editOaut}{\mathsf{editO_{\calA_\varphi}}}

\newcommand{\selEditI}{\mathsf{{sel}\mbox{-} editI_{\varphi_{I}}}}
\newcommand{\selEditO}{\mathsf{{sel}\mbox{-} editO_{\varphi}}}
\newcommand{\selEditIaut}{\mathsf{{sel}\mbox{-} editI_{\calA_{\varphi_{I}}}}}
\newcommand{\selEditOaut}{\mathsf{{sel}\mbox{-} editO_{\calA_\varphi}}}

\newcommand{\minEditI}{\mathsf{{minD}\mbox{-} editI_{\varphi_{I}}}}
\newcommand{\minEditO}{\mathsf{{minD}\mbox{-} editO_{\varphi}}}
\newcommand{\minEditIaut}{\mathsf{{minD}\mbox{-} editI_{\calA_{\varphi_{I}}}}}
\newcommand{\minEditOaut}{\mathsf{{minD}\mbox{-} editO_{\calA_\varphi}}}

\newcommand{\readInp}{\mathsf{read\_in\_chan}}
\newcommand{\readOut}{\mathsf{read\_out\_chans}}
\newcommand{\release}{\mathsf{release}}

\newcommand{\sft}[1]{\textup{\sffamily{#1}}}

\newcommand{\squishlist}{
	\begin{list}{$\bullet$}
		{ \setlength{\itemsep}{0pt}
			\setlength{\parsep}{1pt}
			\setlength{\topsep}{1pt}
			\setlength{\partopsep}{0pt}
			\setlength{\leftmargin}{0.9em}
			\setlength{\labelwidth}{1.5em}
			\setlength{\labelsep}{0.4em} } }
	\newcommand{\squishend}{
\end{list}  }

\newtheorem{proposition}{Proposition}
\begin{document}

\title{Runtime Interchange for Adaptive Re-use of Intelligent Cyber-Physical System Controllers}

\author{Hammond Pearce, \emph{Member, IEEE};
		Xin Yang;
		Srinivas Pinisetty; 
		Partha S. Roop, \emph{Member, IEEE}%
\thanks{H. Pearce was with the Department
of Electrical, Computer, and Software Engineering, University of Auckland, Auckland, New Zealand. E-mail: hammond.pearce@auckland.ac.nz.}%
\thanks{Xin Yang is with the University of Auckland, Auckland, New Zealand.}%
\thanks{S. Pinisetty is with IIT Bhubaneswar, Jatni, Odisha, India.}%
\thanks{P. S. Roop is with the University of Auckland, Auckland, New Zealand.}
\thanks{Manuscript received xxxxxx; revised xxxx.}}
\maketitle

\begin{abstract}
	Cyber-Physical Systems (CPSs) such as those found within autonomous vehicles are increasingly adopting Artificial Neural Network (ANN)-based controllers. 
	To ensure the safety of these controllers, there is a spate of recent activity to formally verify the ANN-based designs. 
	There are two challenges with these approaches: (1) The verification of such systems is difficult and time consuming. 
	(2) These verified controllers are not able to adapt to frequent requirements changes, which are typical in situations like autonomous driving. 
	This raises the question: how can trained and
        verified controllers, which have gone through expensive training
        and verification processes, be re-used to deal with requirement
        changes?

This paper addresses this challenge for the first
time by proposing a new framework that is
capable of dealing with requirement changes at runtime through a mechanism we term \textit{runtime interchange}.
 Our approach functions via a continual exchange and selection process of multiple pre-verified controllers. 
	It represents a key step on the way to component-oriented engineering for intelligent designs, as it preserves the behaviours of the original controllers while introducing additional functionality.
	To demonstrate the efficacy of our approach we utilise an existing autonomous driving case study as well as a set of smaller benchmarks.
	These show that introduced overheads are extremely minimal and that the approach is very scalable.
	
\end{abstract}


\IEEEpeerreviewmaketitle

\section{Introduction}

\acfp{CPS}~\cite{alur2015principles} intertwine distributed controllers,
which are used for controlling physical processes. Traditionally, these used
model-driven approaches to ensure safety. The
model driven approaches are no longer suitable with the introduction of tasks
such as machine vision in industrial \ac{CPS} and
autonomous vehicles. These tasks are primarily handled by machine
intelligence functionality programmed using data-driven methods such
as \acp{ANN}. This is leading to the need for new system design
paradigm where model-driven and data-driven approaches need to be
considered in an unified fashion~\cite{tripakis2018data,Ding2016Modeling}. 
 
Considering this, several approaches inspired by formal
methods~\cite{FormalMethods} have been developed for the verification of safety
properties of \acp{ANN}. Seshia et al.~\cite{seshia2016towards} discuss some of
the challenges for such formal analysis. There have been several
attempts at developing scalable formal verification techniques such as
$AI^{2}$~\cite{ai2}. Also, there have been attempts at verified autonomy~\cite{verisig}
using a neural network controller of an autonomous toy racing car
called F1/10. This feed-forward neural network controller is mapped to hybrid
automata~\cite{ivanovCaseHSCC}. This is then analysed by the \emph{Flow*} model checker to ensure
that the vehicle will not collide with the adjoining obstacles, which
are static. 

These attempts at formally verified controllers are laudable for
ensuring safety. However, some key challenges remain. Specifically, both the
training of complex \acp{ANN} and the subsequent static verification of these controllers are time
consuming processes with scalability concerns.
Finally, it is becoming clear that complex CPS such as autonomous vehicles require \textit{adaptive} controllers which can cope with requirements changes at runtime (see the recent survey \cite{Tavcar2019Review}).
\emph{This poses the question: how to re-use trained and
verified \ac{ANN}-based controllers, which have been developed with
considerable effort, so as to deal with frequent requirement
changes?} 

\ignore{
\acfp{ANN} have excelled in many applications that range from
image and video processing to pattern classification. 
With the advent
of deep neural networks, they have excelled at achieving a level of
precision that far exceed humans while solving
repetitive problems, particularly those related to machine
vision.  
As such, they are beginning to be adopted for use within complex \acfp{CPS} such as industrial systems and autonomous vehicles.
\acp{CPS} such as these consist of one or more distributed embedded systems (termed `the controllers'), which are then used
for controlling physical processes (termed `the plant').

Traditionally, \acp{CPS} are designed using formal models~\cite{FormalMethods} which can be analysed and validated at a
high level of abstraction prior to exposing the validated models to
automated code generation over diverse targets. 
These models are usually designed to be modular and reusable, allowing for libraries of components to be created as design resources.
Such approaches for \ac{CPS} are classified as \emph{model-based} or \emph{component-oriented} design. 
There are robust tools
such as SCADE~\cite{SCADE} based on these philosophies which can generate correct-by-construction code compliant with safety standards.

However, typical \ac{ANN} design strategies are largely not compatible with a model-based approach.
Instead, neural network controllers are designed using \emph{data-dominated} techniques, for instance using labelled images to train deep \acfp{CNN} to recognise and classify incoming camera streams~\cite{tripakis2018data}. 
For safety-critical systems such as autonomous vehicles, verifying these networks may be a time-consuming and difficult process~\cite{TowardsScalableVerificationForDNNS}.
Furthermore, it tends to result in neural networks which are tightly coupled with their implementation and environment, and any change to their systems requires duplication of the training and validation process lest there be unexpected operational consequences.

Considering this, there is recent research momentum in achieving convergence between the usual
model-driven approaches used in \ac{CPS} and the more recent data-driven
approaches used in conventional \ac{AI} based on statistical techniques~\cite{tripakis2018data}.
In~\cite{seshia2016towards} Seshia et al. consider some key challenges of 
using formal methods to verify \acp{ANN}. These include the
difficulty in creating mathematical models and the difficulty in
formalising requirements. There is also the
added challenge of designing scalable verification algorithms for
static analysis of neural networks. More recently, there has been
an attempt at developing abstract interpretation-based
solution~\cite{Gehr2018AI2SA} for the verification of CNNs. 

However, these model-based approaches focus on using modelling to solve the problems inherent in the difficult validation and verification domains, leaving data-driven design pathways for systems involving \acp{ANN}.
We instead propose a different approach. 
The expenses of the time-consuming and resource-intensive validation process can be mitigated if the final product is \emph{re-usable}, which is why many model-driven approaches also focus on component re-usability at their core.
Yet, the model-driven approach for neural networks have not emphasised this, and to our knowledge there are no documented formal approaches for using neural networks for applications outside their original scope.

\red{CPS applications such as autonomous vehicles need real-time decision
making. This not only requires functional but also timing
validation. {\bf To the best of our knowledge,
	there are no known methods for the systematic design of 
	AI applications with timing requirements. }}
}

Here we present a formal methodology to address this issue. 
Our work falls into \textit{runtime assurance}~\cite{RuntimeAssuranceForComplexCPS}, where system behaviours are augmented at runtime by supervisory systems.
The purpose of our framework, which we term ``\acf{RI}'', is to
provide an environment where re-usable controller components can be
composed and interchanged at runtime
 in order to achieve goals outside each component's original
 scope. \ac{RI} is capable of reusing controllers developed using
 traditional model-driven approach with new types of data-driven
 controllers such as neural networks. We motivate the approach using
 the following case study.

\subsection{Motivating Example: F1/10 autonomous vehicle}
\label{sec:motivating}
\begin{figure}[b!]
	\centering
	\vspace{-5mm}
	\includegraphics[width=0.25\textwidth]{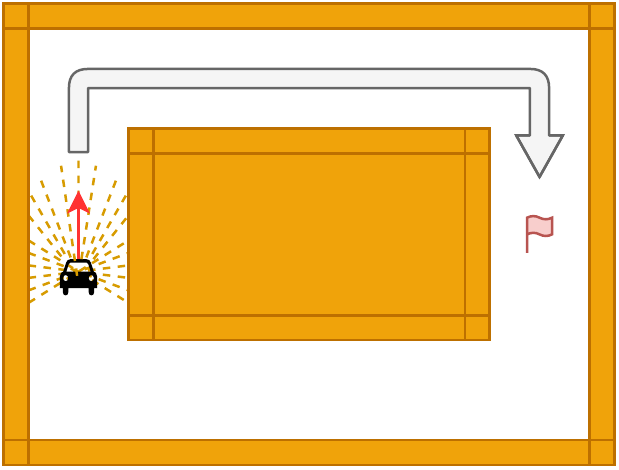}
	\caption{Autonomous F1/10 driving course}
	\label{fig:originalmap}
\end{figure}

We use as an example an F1/10 racing car that drives around a square
track (Figure~\ref{fig:originalmap}). 
The car is controlled using an \ac{MLP} (a type of neural network) trained using LiDAR data to avoid crashing~\cite{ivanovCaseHSCC}.

\begin{figure}[t!]
	\centering
	\includegraphics[width=0.25\textwidth]{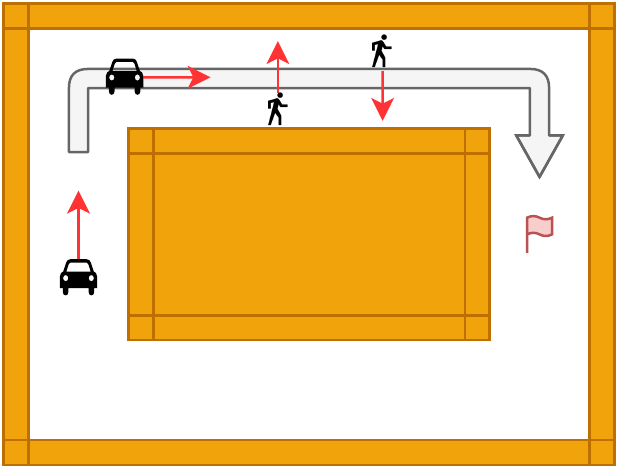}
	\caption{Other vehicles and pedestrians on the driving course}
	\label{fig:newmap}
	\vspace{-6mm}
\end{figure}

This original verified controller only manages the steering of the vehicle and is trained only to avoid collision with the walls.
However, real-world environments are often chaotic and unpredictable.
Consider the case where a bystander unexpectedly walks onto the car's track, or the case where a second car attempts to follow behind the first car.
These are not a contingencies that the autonomous network was trained for, and as such it will not react appropriately, and the following car may crash into the leading car.
We depict these two situations as Figure~\ref{fig:newmap}.

\ignore{
	\subsection{The original F1/10 System}
	
	The rectangular racing track from \cite{ivanov2019case} is depicted in Figure~\ref{fig:originalmap}.
	It is $20m$ by $10m$ with a width of $1.5m$.
	The racing car starts from the mid-point of the left side of the track, drives forwards and makes two left turns, and will finally stop at the mid-point of the right slide of the track. 
	This takes $20s$.

	The vehicle's initial speed is $0m/s$, and it accelerates according to the following equation:
	\begin{equation}
	a = A_c \cdot M_c \cdot (T_c - H_c) - A_c \cdot V
	\end{equation}
	where,
	\begin{itemize}
		\item $a$ represents the vehicle's acceleration
		\item $A_c$ represents the vehicle's acceleration constant, default value is $1.633$
		\item $M_c$ represents the vehicle's motor constant, default value is $0.2$
		\item $T_c$ represents the vehicle's throttle constant, default value is $16$
		\item $H_c$ represents the vehicle's hysteresis constant, default value is $4$
		\item $V$ represents the vehicle's current velocity
	\end{itemize}
	
	Hence, the vehicle will keep accelerating until its velocity reaches $0.2\times(16-4)=2.4m/s$.
	
	The vehicle has a LiDAR with $230$ degrees of coverage, with approximately one ray per $3.8$ degrees.
	In the original controller, the length of these rays are fed into the vehicle's \ac{ANN} as a group of inputs. 
	The \ac{ANN} controller will then produce a desired steering control for the vehicle.
	In the original implementation, the \ac{MLP} is trained in this map using reinforcement learning to avoid collisions with the walls.
	To quantify a collision, the car's shape is abstracted as a circle with a radius of $0.3m$.
	If this circle touches one of the surrounding walls this it is considered a collision.
}

\begin{figure}[b!]
	\vspace{-5mm}
	\centering
	\includegraphics[scale=0.37]{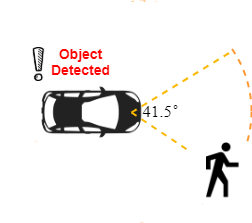}
	\caption{Front LiDAR}
	\label{fig:frontlidar}
\end{figure}

In these scenarios it is possible that the original \ac{MLP} may still be suitable to control the steering of the autonomous cars, as it was designed to handle noisy data.
However, additional functionality is required to control the throttle of the vehicle.
Intuitively, if an obstacle is detected, the vehicle needs to apply brakes and slow down to avoid collision. 
Then, if the obstacle leaves, the vehicle can return to its original speed. 
In order to detect the obstacles, we can utilise the existing LiDAR, scanning for the minimum distance in the central $41.5$ degrees of each scan, as represented in Figure~\ref{fig:frontlidar}.

\ignore{The LiDAR and speedometer can also be used when considering safe autonomous behaviour in the presence of other vehicles.
For instance, the same detection method presented in  Figure~\ref{fig:frontlidar} could be used to detect other vehicles on the track ahead.
However, the minimum distance reading of the LiDAR will not change in the same way as it would for a pedestrian, as the other car will continue to drive around the track as well.
As such, simply braking to a stop (the behaviour invoked for a pedestrian) is not suitable.
Instead, a PI-based car following controller~\cite{xiao2011practical} should be enabled so that the other car can be trailed at a safe distance.}

This leads us to the main contributions of the paper. (1) We present a
formal approach, called \acf{RI}, for combining pre-verified controllers developed
using model-driven and data-driven approaches. RI
significantly extends existing techniques based on runtime
enforcement of reactive systems~\cite{de2018safety, RuntimeEnforcementOfCPS}, which deal only with
single controllers. 
(2) We present a demonstration of our system, analysing scalability and overhead, and demonstrate the enforcement of timed policies to deal with runtime requirement changes for the first time (to the best of
our knowledge). 

The rest of the paper is organised as follows: 
\S\,\ref{sec:rex} introduces RI. 
\S\,\ref{sec:synthesis} then presents our methodology for automatically synthesizing our \ac{RI} code. 
\S\,\ref{sec:results} then presents benchmarking results.
Finally, \S\,\ref{sec:related} presents an overview of the related work in this area, and \S\,\ref{sec:conclusions} concludes the
paper.

\section{Runtime Interchange for Intelligent Controllers}
\label{sec:rex}

Three distinct modes are necessary to meet the runtime requirement changes in the F1/10 course in Figure~\ref{fig:newmap} --- \emph{normal}, where the car is driving as in the original case study \cite{ivanovCaseHSCC}, 
\emph{stopping}, where the car must stop in reaction to an object on the track, and \emph{cautious}, which is where the car needs to slow down either to follow the kinematics of a vehicle in front of it or in preparation for stopping for an object on the track.

Each of these modes will have groups of controllers that are specialised for their operation.
For instance, we adopt the \ac{MLP} steering controller from~\cite{ivanovCaseHSCC} as well as their ``full throttle'' speed controller for the normal mode.
However, if a pedestrian walks onto the road,  different controllers, such as a swerving heuristic as well as a linear braking model are adopted.

Controllers may be utilised by more than one group --- for instance, the steering \ac{MLP} is used in both the \emph{normal} and \emph{cautious} modes.
In the following section we formalise how we can define the limits of these controllers, compose them, and safely switch between them at runtime. 

\subsection{Overview of \acf{RI}}

Inspired by synchronous programming languages~\cite{SynchronousLanguages12YearsLater}, we consider compositions of CPS controllers (including \acp{ANN}) as modular black-box components which run in iterations called  \emph{ticks}. 
During each tick, the overall system first reads sensory input, then processes the inputs using one or more controllers, then finally emits outputs before restarting the loop.

Each of these controller components are structured into \emph{groups}, and each group has a \emph{policy} defining the situations where it may be used.
For instance, the \emph{normal} driving group composes  the ``Steering \ac{MLP}'' with a constant full throttle, whereas the \emph{stopping} driving group combines the ``Swerving Heuristic'' and ``Linear Braking'' controllers.
These groups are then associated with policies, defining when they may be used.

\begin{figure}[h!]
	\center
	\includegraphics[width=0.43\textwidth]{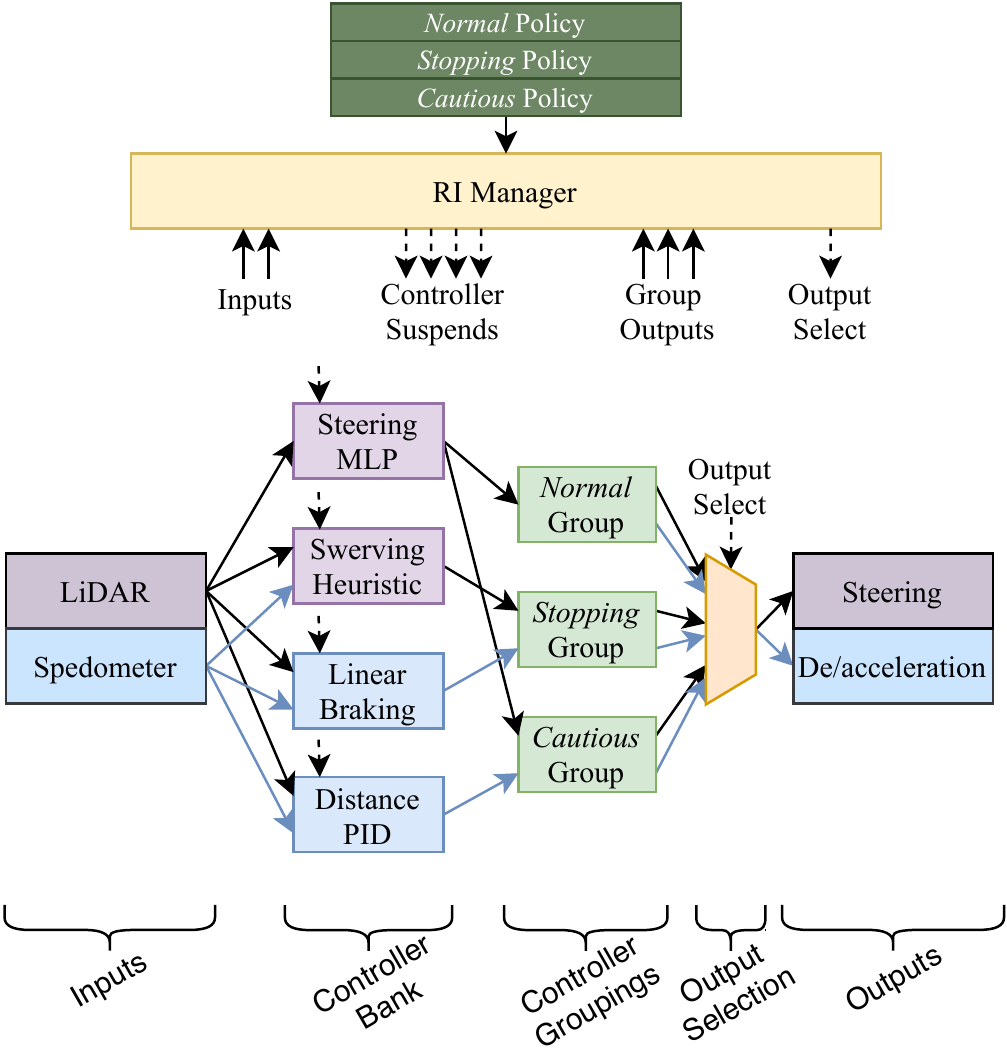}
	\caption{Overall F1/10 \ac{RI} system diagram}
	\label{fig:ri-f110}
\end{figure}

The overall system diagram for \ac{RI} is depicted in   Figure~\ref{fig:ri-f110}. Here, the \ac{RI} Manager is an automatically generated component responsible for the interchange of the different controller groups at runtime.
It does this based upon the aforementioned policies using  \emph{controller suspension}~\cite{SynchronousLanguages12YearsLater}, a technique adopted from synchronous languages.
Suspension freezes a stateful-controller's state when it should not advance, for instance if it would receive data not suitable for it.
In our case, the only controller that would need suspension is the ``Distance PID'', as if it is not currently reading a distance to a car in-front the internal state of the PID may become corrupt.

\ignore{
\subsection{Formalising the network}

In order to specify when our controller groups can be executed, we adopt semantics of \acf{RV}. 
That is, each controller group is associated with a contract.
During normal operation, input output traces are monitored and compared to the specified contracts.
The general operational rule is that if a given input/output would take a contract into a non-accepting state, we reject it by switching to a different contract and controller group.

Let's now examine this more formally.
Our approach considers the controller components $c \in C$ as reactive modules composed into controller groups $f \in F$, where each group $f$ is associated with a contract $\varphi$.
Reactive means they operate in a synchronous loop with their environment.
They first read environmental inputs, then perform some processing, and then emit their outputs before repeating the process.

Overall, we consider the network by the following definition:
\begin{definition}
	\label{def:ri-controller}
	A \ac{RI} network $R$ can be formalised as a tuple $R = \left( I, O, C, F, \right)$, where:
	\begin{itemize}
		\item $I$ is a finite collection of $n$ input variables with
		its domain being $\mathbf{I} =\mathbb{R}^n$,
		\item  $O$ is a finite collection of $m$ output variables with
		its domain being $\mathbf{O} = \mathbb{R}^m$,
		\item $C$ is a finite collection of $o$ controller components $c$, where each $c = \left( I^C, O^C, N, \lambda^C \right)$ such that \squishlist
			\item[-] $I^C \subseteq I$ is a collection of $j$ inputs to the controller component,
			\item[-] $O^C$ is the collection of $k$ outputs from the controller component with its domain being $\mathbf{O^C} = \mathbb{R}^k$
			\item[-] $N$ is a collection of $l$ internal variables inside the controller component with
			its domain being $\mathbf{N} = \mathbb{R}^l$,
			and 
			\item[-] $\lambda^C: I^C \times N  \rightarrow O^C \times N$ is the non-linear
			function (termed the control function) that provides the behaviour of a given controller, i.e.
			when provided vectors of input and internal variables, produces vectors of output variables and updated internal variables,			
		\squishend  
		\item $F$ is a finite collection of $p$ controller groups $f$, where each $f = \left( I^F, O^F, \lambda^F, \varphi \right)$ such that \squishlist
			\item[-]$I^F \subseteq \bigcup O^C, \forall O^C \in C$ is a collection of inputs to the group taken from possible outputs $O^C$ amongst all the controller components $C$ \todo{not sure correctness of formula},
			\item[-] $O^F = O$ is the collection of outputs from the controller group (which intuitively must be the same as the output set for the overall \ac{RI} network),
			\item[-] $\lambda^F: I^F \rightarrow O^F$ is the stateless update function which formats the collection of controller outputs as the collection of network outputs, and
			\item[-] $\mathcal{A}$ is the contract defining when this controller group can be used, which we shall detail later in Definition~\ref{def:vdta}.
		\squishend
		
	\end{itemize}
\end{definition}

\begin{example}
	\label{eg:ri-system}
	Consider the \ac{RI} system as depicted in Figure~\ref{fig:ri-f110}.
	We can formalise the network $R = \left( I, O, C, F \right) $ using Definition~\ref{def:ri-controller} as follows.
	Inputs $I = \{R, v\}$ where $R$ is an array of the current LiDAR distance readings and $v$ represents the velocity of the car as captured by a spedometer.
	The outputs $O = \{d, a\}$ where $d$ represents the current steering command, and $a$ represents the current de/acceleration command.
	There are five controllers in $C = \{c_{sm}, c_{sh}, c_{lb}, c_{fp} \}$ for each of $\{$ ``Steering MLP'', ``Swerving Heuristic'', ``Linear Braking'', ``Follower PID'' $\}$.
	Finally, there are three controller groups in $F = \{f_{norm}, f_{ped}, f_{follow}\}$ for each of $\{$ ``normal'', ``pedestrian'', ``following'' $\}$.
	
	Now let us examine one of the controllers.
	The ``Steering MLP'' $c_{sm} = \langle I^C, O^C, N, \lambda^C \rangle $ can be defined with inputs $I^C = \{R\}$ and outputs $O^C = \{d\}$.
	This makes intuitive sense, as this component simply maps incoming LiDAR data to steering commands without consideration for the speed of the vehicle.
	As the network is a simple feed-forward \ac{MLP}, $N = \emptyset$ (as \acp{MLP} have no internal state), and $\lambda^C$ represents the non-linear update function of the network.
	
	Let's now examine a controller group.
	We can see that the ``normal'' group $f_{norm} = \{I^F, O^F, \lambda^F, \varphi\}$ is made up as follows.
	Firstly, inputs $I^F = \{ \{d\} \in c_{sm}\}$. 
	This means that the inputs to this controller group are just the steering command $d$ coming from the ``Steering MLP'' component.
	Secondly, outputs $O^F = \{d, a\}$, as the outputs of all control groups must match the outputs of the overall system.
	Finally, the internal function $\lambda^F \defeq O^F = \{d\in I^F, 1\}$, i.e. the steering command $d$ from the ``Steering MLP'' is concatenated with a full throttle ($a = 1$) command.
	A contract $\mathcal{A}$ may then be defined for $f_{norm}$, which we shall examine further in Example~\ref{eg:vdta-normal}.	
\end{example}

We then define an execution of the network components as follows:
\begin{definition}
	\label{def:cps-reaction}
	Given an \ac{RI} system $N = \left( I, O, C, F \right)$, we
	define a reaction based on $N$ to be a periodic invocation of $N$.
	This involves invocation of all \emph{non-suspended} internal controllers in $C$ followed by invocation of all controller groups in $F$.
	For each controller $c \in C$, an invocation refers to the triggering of the internal update function $\lambda^C$.
	Then, for each controller $f \in F$, an invocation refers to the triggering of the internal update function $\lambda^F$.
	Each of these invocations requires one \emph{tick} to complete, where a tick
	denotes the fixed period of a logical clock.
\end{definition}

Let us now consider how contracts may be specified, as these will detail how we manage the overall \ac{RI} system at runtime.
}

\subsection{Preliminaries and Notations for Runtime Interchange}
\label{sec:prelim}

We consider Intelligent Cyber-Physical Systems to have finite ordered sets of valued input channels ${I} = \{{i_1}, {i_2}, \ldots {i_n}\}$ and valued output channels ${O} = \{{o_1}, {o_2}, \ldots {o_n}\}$.
For a variable (resp. channel) $v$, ${\mathcal D}_v$ denotes its domain,
and for a finite ordered set of variables $V= \{v_1, \ldots, v_n \big\}$,
${\mathcal D}_V$ is the product domain ${\mathcal D}_{v_1} \times \cdots \times {\mathcal D}_{v_n}$. 

Consider $n \in \bbn$, $\bbb_n$ denotes the domain of the finite ordered set of Boolean $\{b_1, \cdots, b_n\}$.
A valuation of the variables in $V$
is a mapping $\nu$ which maps every variable $v \in V$ to a value $\nu\big(v\big)$ in ${\mathcal D}_v$.

A finite (resp. infinite) word over $\calD_C$ (where $C = I \cup O$) is a finite sequence $\sigma = \eta_1\cdot \eta_2 \cdots \eta_n$ where $\forall i \in [1,n]:$ $\eta_i$ is a tuple of values of variables in $C = I \cup O$. For convenience where necessary, each element $\eta_i$ is considered to be a pair $\left(\eta_I, \eta_O\right)$, where $\eta_I$ is a valuation of all the variables in $I$, and   $\eta_O$ is a valuation of all the variables in $O$.
The set of finite (resp. infinite) words over $\calD_C$ is denoted by $\calD_C^*$ (resp. $\calD_C^\omega$).
The {\em length} of a finite word $\sigma$ is $n$, denoted $|\sigma|$.
The empty word over $\calD_C$ is denoted by $\epsilon_C$, or $\epsilon$ when clear from the context.
$\calD_C^+$ denotes $\calD_C^*\setminus\{\epsilon\}$.
The {\em concatenation} of two words $\sigma$ and $\sigma'$ is denoted by $\sigma\cdot \sigma'$.
A word $\sigma'$ is a {\em prefix} of a word $\sigma$, denoted as $\sigma' \pref \sigma$, whenever there exists a word $\sigma''$ such that $\sigma = \sigma'\cdot \sigma''$; conversely $\sigma$ is said to be an \emph{extension} of $\sigma'$.

Given an input-output word $\sigma= (x_1,y_1)\cdot(x_2,y_2)\cdots(x_n,y_n) \in \calD_C^*$, the input word obtained from $\sigma$ is denoted by $\sigma_I$ where $\sigma_I = x_1 \cdot x_2 \cdots x_n \in \calD_I^*$ is the projection on inputs ignoring outputs.
Similarly, the output word obtained from $\sigma$ is denoted by $\sigma_O$ where $\sigma_O = y_1 \cdot y_2 \cdots y_n \in \calD_O^*$ is the projection on outputs.

Given a word $\sigma$ and $i \in [1, |\sigma|]$, $\sigma_{[i]}$ denotes the element at index $i$ in $\sigma$. Given a word  $\sigma$ and two integers $i, j \text{ s.t. } 1 \leq i \leq j \leq  |\sigma|$, the subword of $\sigma$ from index $i$ to $j$ is denoted as $\sigma_{[i\cdots j]}$. 
Given an n-tuple of symbols $e = (e_1, \cdots, e_n)$, for $i \in [ 1,n]$, $\pi_i (e) = e_i$ denotes the projection of $e$ on its $i$-th element. 
The operator $\pi_i$ is naturally extended to words of n-tuples of symbols to produce the word formed by the concatenation of the projections on the $i^{th}$ element of each tuple.

In every tick, the RI manager first examines the input from the environment, and later the output from the controller.
The overall output of the RI manager in every tick is an input-output event. 
We introduce function IO which is used to treat an input word $\sigma_I \in \calD^*_I$, and an output word $\sigma_O \in \calD^*_O$ as in input-output word in $(\calD_I \times \calD_O)^*$.

Function IO:
Given an input word $\sigma_I = x_1 \cdot x_2 \cdots x_n \in \calD_I^*$ and an output word 
$\sigma_O = y_1 \cdot y_2 \cdots y_n \in \calD_O^*$ s.t. $|\sigma_i| == |\sigma_o|$,
$IO(\sigma_i, \sigma_o) = (x_1, y_1) \cdot (x_2, y_2) \cdots (x_n, y_n)$.

\ignore{
	Function $IO$: Given a word $\sigma_I \in \calD_I^*$ and a word $\sigma_o \in \calD_O^*$ s.t. $|\sigma_i| == |\sigma_o|$, $IO(\sigma_i, \sigma_o) = \sigma $, where $\sigma \in \calD_C^*$ and $|\sigma| ==|\sigma_i|$ and $\forall j \in [1, |\sigma|], \pi_1(\sigma_{[j]})== \sigma_i{_{[j]}})$, and $\forall j \in [1, |\sigma|], \pi_2(\sigma_{[j]})== \sigma_o{_{[j]}})$. 
}

A property $\varphi$ over $C$ defines a set $\calL\left(\varphi\right)\subseteq \calD_C^{*}$.
A program $\calP \models \varphi$ iff $\calL\left(\calP\right) \subseteq  \calL\left(\varphi\right)$.
In this paper, properties are formally defined as VDTA.

\subsection{Policy Specification}

\ignore{
A \ac{VDTA}
constrains the values and the timing  of inputs / output events in an industrial control system.
\ac{VDTA} is used in our setting as a distinct contract for each controller group $f$ inside the \ac{RI} system. 
These will define the operating boundaries of each group while
allowing the selection between controller groups at runtime, as the
\ac{RI} manager decides which controllers should execute and which
need to be suspended. Before we examine the formal definition of \ac{VDTA}, let us consider an example.}

We adopt \acf{VDTA}, a language for expressing policies over industrial and cyber-physical
 systems~\cite{SmartIOModules}.
A \ac{VDTA} can be seen as an automaton with a finite set of locations, a
finite set of discrete clocks used to represent time evolution, and
external input (resp. output channels) called ``external variables''
which are used for representing system data. They model the data from
the monitored system (resp. environment) read from the input (resp.)
channels in every tick. 
In a \ac{VDTA}, time evolves synchronously: that is, the system
executes as a series of discrete \emph{logical ticks} 
where each tick takes exactly one transition~\cite{SynchronousLanguages12YearsLater}.
In the semantics of VDTA, each transition will be associated with
values of external variables.

\begin{example}
	\label{eg:vdta-norm}

	Let us consider the required behaviour for the \ac{RI} system as depicted in Figure~\ref{fig:ri-f110} for the autonomous vehicle.
	In order to safely move through the environment in Figure~\ref{fig:newmap}, this system describes three control groups (modes): \emph{normal}, \emph{stopping}, and \emph{cautious}.
	Each of these modes will need an associated policy to describe when it should and should not be used. 
	If one mode is active, and it becomes unsuitable for use, the system should automatically change to another mode.
	
	Firstly, let us examine the \emph{normal} mode. 
	This represents the original control scheme from \cite{ivanovCaseHSCC}, and is made from just the ``Steering MLP'' controller, with the acceleration control simply set to maximum at all times (taking the car to its maximum speed).  
	This mode can be considered safe for use until an object is detected on the track in front of the vehicle. 
	To test for this, we define a function $\sft{min}(R)$ which returns the minimum distance in the middle 41.5$^o$ of the LiDAR rays (e.g. Figure~\ref{fig:frontlidar}).
	If $\sft{min}(R)$ ever returns a distance less than or equal to 1.2 units, we say this is ``unsafe'', and this group is no longer suitable for use. 
	In addition, for it to become ``safe'' again, we require six consecutive readings where $\sft{min}(R)$ is greater than 1.2 units.
	This requirement prevents sensor noise accidentally changing our system from ``unsafe'' to ``safe''.
	
	\begin{figure}[b!]
		\vspace{-5mm}
		\center
		\begin{subfigure}{0.48\textwidth}
			\centering
			\includegraphics[scale=0.8]{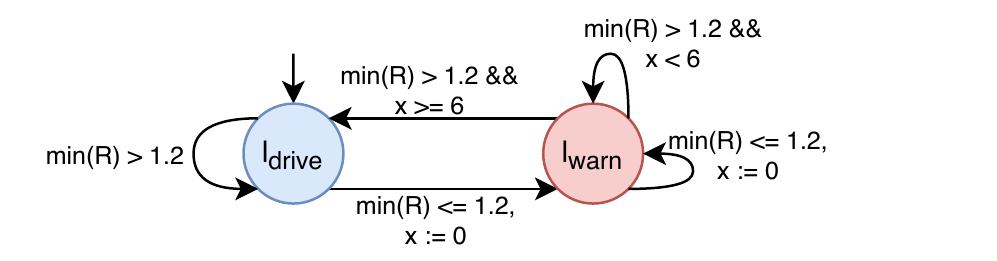}
			\caption{Policy for \emph{normal} mode group $\mathcal{A}_{norm}$}
			\label{fig:vdta-norm}
		\end{subfigure}
		
		\begin{subfigure}{0.48\textwidth}
			\centering
			\includegraphics[scale=0.8]{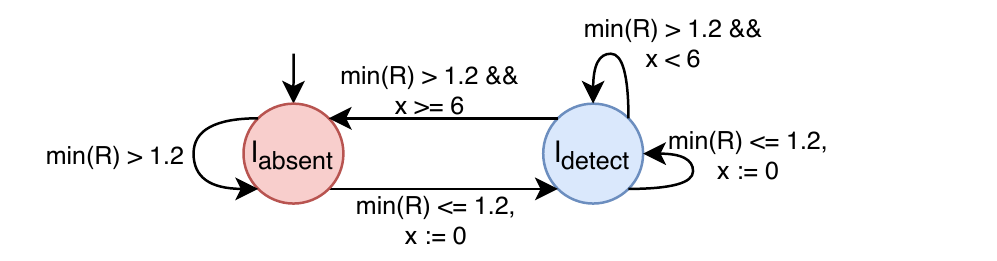}
			\caption{Policy for \emph{stopping} mode group $\mathcal{A}_{stop}$}
			\label{fig:vdta-stop}
		\end{subfigure}
		
		\begin{subfigure}{0.48\textwidth}
			\centering
			\includegraphics[scale=0.8]{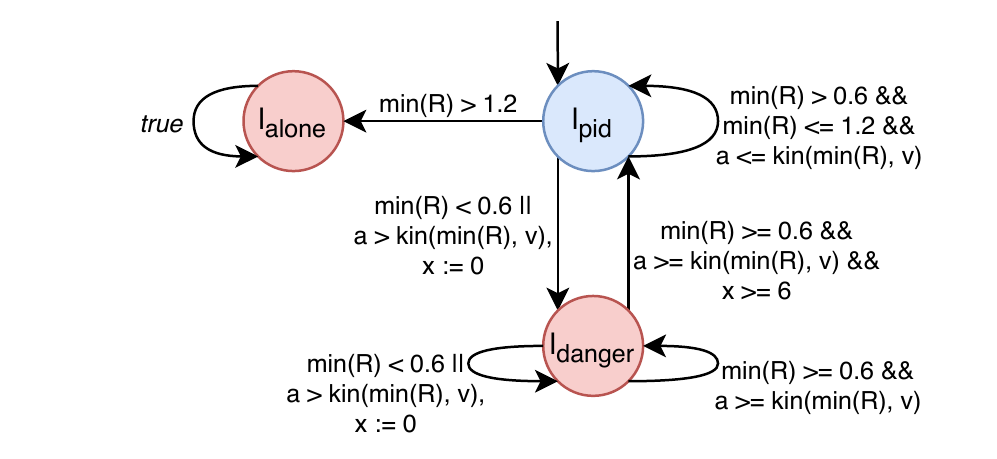}
			\caption{Policy for \emph{cautious} mode group $\mathcal{A}_{caut}$}
			\label{fig:vdta-caut}
		\end{subfigure}
		\caption{Control group policies}
		\label{fig:vdta}
	\end{figure}
	
    We can encode these two rules as the \ac{VDTA} policy $\mathcal{A}_{norm}$ depicted in Figure~\ref{fig:vdta-norm}.
	Here, the policy begins in the ``safe'', accepting location $l_{drive}$, denoted as safe by its blue colour.
	While the policy remains in this location the policy is itself considered accepting, meaning that while it stays in this location, the associated controller group $f_{norm}$ is accepting (i.e. suitable for use at this time).
	
	The controller group may only stay in this location when the function $\sft{min}(R)$ returns a distance $>1.2$. 
	Should a reading ever be $<= 1.2$, meaning an object is within 1.2 units of the front of the car, the policy moves to the non-accepting $l_{warn}$ location, denoted as non-accepting by its red colouring.
	While the policy remains in this location, the associated controller group is considered non-accepting (i.e. unsuitable for use at this time).
	Note also the timer reset $x := 0$ on the transition from $l_{drive}$ to $l_{warn}$. 
	This sets a timer, $x$, to zero.
	
	There are three options for transitions in this location. 
	Firstly, the distance $\sft{min}(R)$ may continue to return $<=1.2$, which continues to cause timer $x$ to reset and the location to stay set as $l_{warn}$.
	Secondly, the distance $\sft{min}(R)$ may return $>1.2$, meaning that the system could be considered safe.
	However, if timer $x$ is not $>=6$ the self loop transition will be taken, meaning that the policy continues to be considered non-accepting. 
	The self loop transition, as it does not feature a timer reset, will advance the timer.
	Finally, if the distance $\sft{min}(R)$ has returned $>1.2$ six times in a row, the timer $x$ will be $>=6$, and as such, the transition from $l_{warn}$ back to $l_{drive}$ may be taken, indicating that the \emph{normal} mode may be considered accepting once more.
	
	Let us now consider the interaction with other policies.
	The \emph{stopping} mode policy is described in Figure~\ref{fig:vdta-stop} as $\mathcal{A}_{stop}$.
	This policy's control group is the ``Swerving Heuristic''
        paired with ``Linear Braking'', are needed to safely bring the car to a complete halt.
	This policy describes the exact inverse of the \emph{normal} policy. It says that if any reading $\sft{min}(R)$ is less than $1.2$, we move to the accepting location $l_{detect}$ from non-accepting location $l_{absent}$. 
	Like with $\mathcal{A}_{norm}$, if this location observes six consecutive readings where the minimum distance is greater than 1.2 then the policy moves to the original location.
	This would render this policy non-accepting once more.
	
	Intuitively, when $\mathcal{A}_{norm}$ is accepting, $\mathcal{A}_{stop}$ is non-accepting, and vice versa.
	This means that when one of these control modes is suitable for use, the other may be considered unsuitable.
	Should the \emph{normal} mode be active, and a reading $\sft{min}(R)$ be $<=1.2$, then the two policies will switch, and the \ac{RI} Manager should select the outputs from the other control group.
	
	Let us now consider the need for a third control mode.
	While all situations are covered by the first two control modes, and even though is safe to do so in this environment, quality of service is degraded if the car must always come to a halt if any object is detected.
	We mitigate this with the \emph{cautious} control mode $\mathcal{A}_{caut}$, which operates using the ``Distance PID'' controller to manage speed and the ``Steering MLP'' to manage steering of the vehicle.
	To quantify when this control group is suitable for use we must first define a function $\sft{kin}(distance, speed)$ which, given the distance to an object and the current speed of the vehicle returns the minimum safe deceleration to prevent the car crashing into that object. 
	We now define the third mode as accepting when the distance $\sft{min}(R)$ is between 1.2 and 0.6 units.
	If the distance reads lower than 0.6 units we must receive 6 consecutive readings where it is greater than 0.6 before returning to an accepting location.
	We also define that the third mode as non-accepting if it outputs an acceleration $a$ which is not safe according to function $\sft{kin}(distance, speed)$.
	Finally, as large distances have the potential to corrupt the internal state of the PID, we also define that should the distance to the next object $\sft{min}(R)$ be greater than 1.2 units during operation, we do not allow for any return to operation. 
	We encode these rules as as the \ac{VDTA} in Figure~\ref{fig:vdta-caut}. Note the ``trap'' location $l_{alone}$, which has no pathways back to any accepting locations once it has been entered.

\end{example}

\subsection{\ac{VDTA} Syntax and Semantics}
Let $X=\{x_1,\ldots, x_k\}$ be a finite set of integer variables representing discrete clocks.
A {\em valuation} for a clock variable $x$ of $X$  is an element of
$\bbn$, that is a function from $x$ to $\bbn$. 
The set of valuations for the set of clocks $X$ is denoted by $\chi$.
For $\chi\in\bbn^X$, $\chi+1$ (which captures the ticking of the digital clock) is the valuation assigning $\chi\left(x\right)+1$ to each clock variable $x$ of $X$.
Given a set of clock variables $X' \subseteq X$, $\chi\left[X'
  \leftarrow 0\right]$ is the valuation of clock variables $\chi$ where all the clock variables in $X'$ are assigned to $0$.

\begin{definition}[Syntax of \ac{VDTA}s]
	\label{def:vdta}
	An \ac{VDTA} is a tuple \\
	$\calA = \left(L, {l_0}, F, X, I, O, \Delta \right)$ where:
	\squishlist
	\item $L$ is a finite non-empty set of locations, with $l_0 \in L$ the initial location, and $F \subseteq L$ the set of accepting locations;
	\item $X$ is a finite set of discrete clocks;
	\item $I$ is the set of input channels;
	\item $O$ is the set of output channels; 
	\item $\Delta$ is a finite set of transitions, and each transition $t \in \Delta$ is a tuple $\left( l, G, A^X, l' \right)$
	also written\\
	$l \xrightarrow{G\left( I,O \right),A^X} l'$
	such that,
	\squishlist
	\item[\textbullet] $l, l' \in L$ are respectively the origin and target locations of the transition;
	\item[\textbullet] $G = G^D \wedge G^X$ is the guard where
	\squishlist
	\item[-] $G^D = G^I \wedge G^O$ where
	\squishlist
		\item[-] $G^I$ is a computable predicate over inputs i.e., conjunction of constraints of the form $f1\left(I\right) \sharp f2\left(I\right)$, where $f1$ and $f2$ are computable functions over input variables, and 
		$\sharp \in \{ <, \leq, =, \geq, >, \neq \}$;
		\item[-] $G^O$ is a computable predicate over inputs and outputs i.e., conjunction of constraints of the form $f1\left(I \cup O\right) \sharp f2\left(I \cup O\right)$, where $f1$ and $f2$ are computable functions over input and output variables (but requiring at least one of the output variables as an argument), and 
		$\sharp \in \{ <, \leq, =, \geq, >, \neq \}$; 
	\squishend
	\item[-] $G^X$ is a clock constraint over $X$ defined as
          conjunctions of constraints of the form $x \sharp c, x
          \sharp f1\left(I\right), x \sharp f2 \left(I \cup O \right)$
          where $x \in X$ and and $c \in \mathcal{N}$, $f1\left(  I
          \right)$ is a computable predicate over input variables, $f2
          \left(I \cup O \right)$ (requiring at least one of the
          output variables as an argument) is a computable predicate
          over input and output variables, and $\sharp \in \{ <, \leq,
          =, \geq, >, \neq \}$; 
	\squishend
	\item[\textbullet] $A^X \subseteq X$ is the set of clocks to be reset.
	
	\squishend
	\squishend
\end{definition}

\begin{example}	
	\label{eg:vdta-normal-formal}
	The \ac{VDTA} $\mathcal{A}_{norm}$ for Figure~\ref{fig:vdta-norm} has a set of
        locations $L = \{ l_{drive},l_{warn}\}$, with accepting
        locations F = $\{l_{drive}\}$. $l_{drive}$ is also the initial
        location.
	The \ac{VDTA} has the set of input variables $I = \{R, v\}$ and the set of outputs $O = \{d, s\}$.
	
	$\mathcal{A}_{norm}$ features one functions, $min(R)$, introduced in Example~\ref{eg:vdta-norm}.
	
	In an \ac{VDTA}, a transition can have guards involving variables, clocks, and functions.
	For example, the transition from $l_{caution}$ to $l_{drive}$
        happens when both $min(R) < 1.3$ and the value of clock $x > 2$. 
        This implies that there has been three consecutive readings where the minimum distance is greater than $1.3$ units.
        
	Clock values can be reset upon transitions. 
        For example, upon transition from $l_{drive}$ to $l_{caution}$, the value of clock $x$ is reset to 0.

\end{example}

\ignore{	
A finite (resp. infinite) word over $\calD_C$ (where $C = I \cup O$) is a finite sequence $\sigma = \eta_1\cdot \eta_2 \cdots \eta_n$ where $\forall i \in [1,n]:$ $\eta_i$ is a tuple of values of variables in $C = I \cup O$. For convenience where necessary, each element $\eta_i$ is considered to be a pair $\left(\eta_I, \eta_O\right)$, where $\eta_I$ is a valuation of all the variables in $I$, and   $\eta_O$ is a valuation of all the variables in $O$.
The set of finite (resp. infinite) words over $\calD_C$ is denoted by $\calD_C^*$ (resp. $\calD_C^\omega$).
The {\em length} of a finite word $\sigma$ is $n$ and is denoted by $|\sigma|$.
The empty word over $\calD_C$ is denoted by $\epsilon_C$, or $\epsilon$ when clear from the context.
$\calD_C^+$ denotes $\calD_C^*\setminus\{\epsilon\}$.
The {\em concatenation} of two words $\sigma$ and $\sigma'$ is denoted by $\sigma\cdot \sigma'$.
A word $\sigma'$ is a {\em prefix} of a word $\sigma$, denoted as $\sigma' \pref \sigma$, whenever there exists a word $\sigma''$ such that $\sigma = \sigma'\cdot \sigma''$; conversely $\sigma$ is said to be an \emph{extension} of $\sigma'$.
}

\subsubsection{Semantics for \ac{VDTA}}
Let $\calA = \left(L, {l_0}, F, X, I, O, \Delta \right)$  be a \ac{VDTA}.
The semantics of $\calA$ is a timed transition system,
where a state consists of a location, and valuations of clocks $X$.
Each transition is associated with values of external variables in $C$.

\begin{definition}[Semantics of {VDTA}s]
	\label{def:vdta:semantics}
	The semantics of $\calA$ is a timed transition system $\sem{\calA}=\left( Q, q_0, Q_F, \Gamma, \to \right)$, defined as follows:
	\squishlist
	\item $Q = L \times \bbn^X$, is the set of states of the form $q= \left( l,\chi \right)$ where
	$l \in L$ is a location,
	$\chi$ is a valuation of clocks;
	\item $Q_0 = \{ \left( l_0, \chi_{\left[X \leftarrow 0\right]} \right) \}$ is the set of initial states;
	\item $Q_F = F \times  \bbn^X$ is the set of accepting states;
	\item $\Gamma = \{ \eta \mid
	\eta \in {\mathcal D}_{C}  \}$ is the set of transition labels;
	\item $\to\subseteq Q\times \Gamma\times Q$  the transition relation
	is the smallest set of transitions of the form
	$\left( l,\chi \rangle \longrightarrow {\eta} \langle l',\chi'\right)$
	such that  $\exists \left( l, G, A^X, l' \right) \in \Delta$,
	with $G^X\left(\chi + 1, \eta \right) \wedge G^D\left(\eta\right) $ evaluating to {\true},
	and $\chi'=\left(\chi+1\right)[A^X \leftarrow 0]$. 
	\squishend
\end{definition}

A {\em run} $\rho$ of $\sem{\calA}$ from a state $q\in Q$ over a {\em trace} $w =  \eta_1\cdot \eta_2\cdots \eta_n$ is a sequence of moves in $\sem{\calA}$:
$\rho = q \xrightarrow {\eta_1} q_1
\cdots q_{n-1}\xrightarrow {\eta_n} q_{n}$,
for some $n\in\bbn$.
A run is accepted if it starts from the initial state $q_0\in Q$ and ends in an accepted state $q_n \in Q_F$.

\ignore{
	\begin{example}[Run of an \ac{VDTA}]
	\todo{Update this}
	An example run of the EDTA depicted in Figure~\ref{fig:edta-car-rte} is elaborated here.
	Assume $T_{lim} = 5$.
	A run of this EDTA starting from the initial state $\left(l_{safe}, t = 0\right)$ for the word $\sigma = \left(101,0\right)\cdot \left(100,0\right)\cdot \left(90,0.2\right)\cdot \left(101,0.2\right)\cdot\left(75,0.4\right)\cdot \left(70,0\right)$ is:\\
	{\small$
		\left(l_{safe}, t = 0\right)
		\xrightarrow {\left(101, 0\right)} 
		\left(l_{safe}, t = 1\right)
		\xrightarrow {\left(100, 0\right)} 
		\left(l_{brake}, t = 0\right)
		\xrightarrow {\left(90, 0.2\right)} \\
		\left(l_{brake}, t = 1\right)
		\xrightarrow {\left(101, 0.2\right)} 
		\left(l_{brake}, t = 2\right)
		\xrightarrow {\left(75, 0.4\right)} 
		\left(l_{brake}, t = 3\right)
		\xrightarrow {\left(70, 0\right)} 
		\left(l_{vio}, t = 4\right).
		$
	}
	
	The run started in the initial state.
	A pedestrian is detected in the second tick, and in the third tick, the \ac{AV} starts braking with $B = 0.2$. 
	This meets the safe threshold.
	In the fourth tick, the input sensor package misclassifies and says that no pedestrian is detected, setting $P = 101$.
	However, the controller continues to brake with $B = 0.2$.
	The pedestrian is re-detected in the fifth tick and the car continues to brake.
	However, in the sixth tick, the controller malfunctions, and even though a pedestrian is still detected, it stops braking.
	As a result, the \ac{VDTA} will go to the non-accepting state $l_v$. 
	This is thus a non-accepting run and represents a violation scenario.
\end{example}
}
\begin{definition}[Deterministic (complete) \ac{VDTA}]
	\label{def:detComplete}
	A \ac{VDTA} $\calA= \left(L, {l_0}, F, X, I, O, \Delta \right)$ with its semantics $\sem{\calA}$ is said to be a {\em deterministic} \ac{VDTA} whenever for any location $l$
	and any two distinct transitions $\left(l,g_1,A^X_1,l'_1\right) \in \Delta$ and $\left(l, g_2, A^X_2, l'_2 \right)\in \Delta$ with same source $l$, the conjunction of guards $g_1\wedge g_2$ is unsatisfiable.
	$\calA$ is {\em complete} whenever for any location $l\in L$ the disjunction of the guards of the transitions leaving $l$ evaluates to {\em true}.
\end{definition}

\ignore{
\begin{definition}[Product of EDTAs]
	\todo{Srinivas: update}
	\label{EDTA:product}
	Given two EDTAs 
		\[\calA^{1} = \left(L^{1}, {l_0}^{1}, X^{1}, V^{1}, C^{1}, \Theta^{1}, F^{1},  \Delta^{1} \right)\] and
		\[\calA^{2} = \left(L^{2}, {l_0}^{2}, X^{2}, V^{2}, C^{2}, \Theta^{2}, F^{2},  \Delta^{2} \right)\]
		with disjoint sets of integer clocks (X), internal variables (V) and external variables (C), their product is the VDTA \[\calA^{1}\times \calA^{2}= \left(L, {l_0}, X, V, C, \Theta, F,  \Delta \right)\] where
	$L=L^1 \times  L^2$, $l_0 = (l^1_{0}, l^2_{0})$,  $X  = X^1 \cup X^2$  , $V  = V^1 \cup V^2$ ,  $C  = C^1 \cup C^2$, $F = F^1 \times F^2$, 
 $\Theta = \Theta^{1} \wedge \Theta^{2}$
	and  $\Delta$ 
	is a finite set of transitions where each transition $t=\left( \left(l^{1},l^{2}\right), G^{1}\wedge G^{2}, A^{1} \cup A^{2}, \left(l^{'1},l^{'2}\right) \right)$ belongs to $\Delta$ if $\left( l^{1}, G^{1}, A^{1},l^{'1} \right)$ belongs to $\Delta^{1}$ and  $\left(l^{2}, G^{2}, A^{2},l^{'2} \right)$ belongs to $\Delta^{2}$.
\end{definition}

The product of \acp{VDTA} is useful when we have multiple properties to enforce.
}

\ignore{
\subsection{Edit Functions}
\label{sec:editfunc}
The proposed extension mechanism is allowed to augment input (resp. output) channels when necessary.
The augmentation functions that we introduce here will be used in defining the extension and extension algorithm in the later sections. 
}

\subsection{Controller Component Suspension}

While in the simple case all controller components would execute every tick, in reality not all controller designs are suitable to be executed over all possible inputs.
Consider the policy in Figure~\ref{fig:vdta-caut}, with the ``trap'' location $l_{alone}$, representing the situation where the internal state of the ``Distance PID'' may have become corrupted by a reading of $\sft{min}(R)$ being too large.
This is a ``trap'' location as it is a non-accepting location with no possible path leading back to an accepting location.

Due to the presence of the ``trap'' location, we can detect that this scenario will occur should we present the large $\sft{min}(R)$ input to the controller.
As a result, instead of presenting the input, we instead \emph{suspend} the associated policy and controller, freezing their state and preventing corruption.
In other words, if a policy is guaranteed to advance to a trap
location given the input to the \ac{RI} network, we suspend operation
of that policy and its associated controllers. Consequently, the trap
location is never entered in any policy due to the associated
suspension. This is further elaborated using Remark~\ref{remark:suspend}.

To achieve this, we must compare inputs from the environment over control group policies in the absence of controller outputs.
We thus need to consider the input property that we obtain from $\calA_\varphi$ by projecting on inputs~\cite{RuntimeEnforcementOfCPS,SmartIOModules}.

\begin{remark}[Input \ac{VDTA} $\calA_{I}$]
	\label{def:inp:prop:proj:def}
Given a property defined as \ac{VDTA} $\calA=\left(L, {l_0}, F,  X, I, O,  \Delta \right)$, input \ac{VDTA} $\calA_{I}=\left(L, {l_0}, F,  X,  I,  \Delta_I \right)$ is obtained from $\calA$ by ignoring outputs channels on the transitions. 
The structure of the input automaton $\calA_{I}$ that we obtain will be exactly identical to the automaton $\calA$.  
For every transition $t \in \Delta$ there will be a transition $t' \in
\Delta_I$ where $t' = \left( l, G', A^X, l' \right)$ is obtained from
t = $\left( l, G, A^X, l' \right)$ (with $G = G^D \wedge G^X$ where
$G^D = G^I and G^O$), by discarding $G^O$ in $G^D$ and discarding
clock constraints in $G^X$ with the function on the right-hand side
requiring an output channel as an input parameter. Input \acp{VDTA}
may be non-deterministic. 
\end{remark}

Input \ac{VDTA} $\calA_I$  described in remark \ref{def:inp:prop:proj:def} is a property of inputs (only) that we obtain from a \ac{VDTA} $\calA$ that expresses a property over both inputs and outputs. 
Once obtained, we can first compare an input value to the input \ac{VDTA} for a given group. 
If that group would move into a ``trap'' location, i.e. a location which guarantees non-acceptance forever, we suspend the controllers associated with this group and we do not update the policy this tick.
This process is discussed further in the next section. 

\begin{figure}[htb]
	\centering
	\vspace{-7mm}
	\includegraphics[scale=0.8]{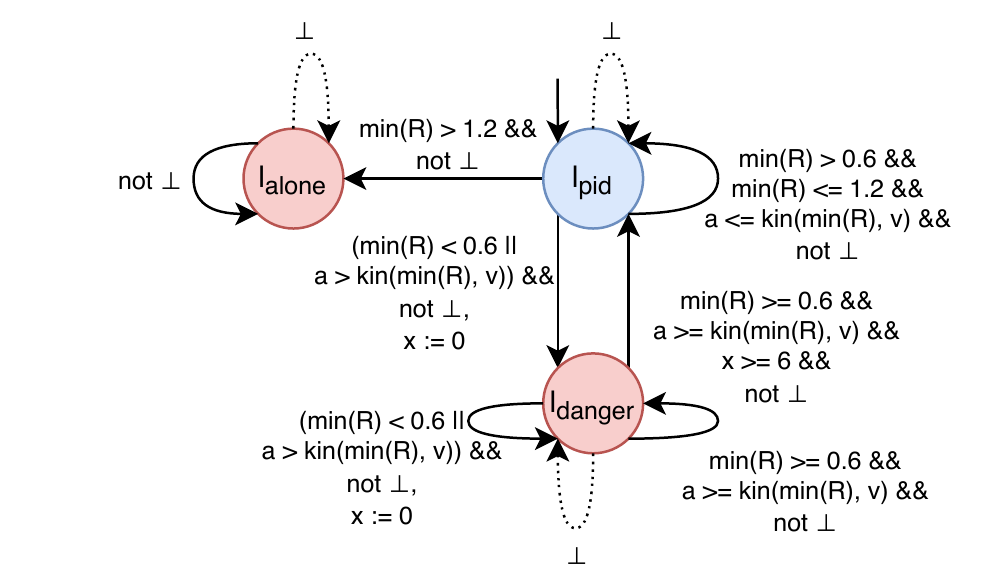}
	\vspace{-2mm}
	\caption{Suspendable policy for \emph{cautious} group $\mathcal{A}_{caut}$}
	\label{fig:vdta-following-suspendable}
	\vspace{-5mm}
\end{figure}

\begin{remark}
\label{remark:suspend}
	As discussed, controller groups are suspended if an input would advance it to a trap location.
	Here we describe the mechanism for this.
	Let $\varphi_i$ be the policy (defined as $\calA_{\varphi_i}$) corresponding to the  controller group $i$. 
	Let $x \in \calD_I$ denote the input in a particular tick. 
	\begin{itemize}
		\item If the controller group $i$ will not be suspended in that particular tick, the the state of the policy $\calA_{\varphi_i}$ will be updated by consuming the event $(x,y) \in \calD_C$ where $y$ is the output produced by that controller group in that tick.
		\item If the controller group $i$ will be suspended in
                  a given tick, for the updation of the policy $\calA_{\varphi_i}$ w.r.t. the input $x \in \calD_I$ that is observed in that tick, the output of the controller group $i$ in that tick is considered to be invalid/empty (denoted as $\bot$). For every location in $\calA_{\varphi_i}$, for all $x \in \calD_I$ we consider implicit self transitions with event $(x, \bot)$, allowing the policy to remain in the same location when the controller group is suspended.      
	\end{itemize}
Thus, the output domain $\calD_O$ will be will be considered as $\calD_O \cup  \{\bot\}$. 
An example of this is presented in Figure~\ref{fig:vdta-following-suspendable}. If the policy is in location $l_{pid}$, and $\sft{min}(R) > 1.2$, then the \ac{RI} Manager emits $\bot$ for this policy and suspends the associated ``Distance PID'' controller.
As a result, the self loop in $l_{pid}$ is taken instead of advancing to location $l_{alone}$.

\end{remark}

\ignore{

\begin{figure}[tb]
	\centering
	\includegraphics[width=0.5\textwidth]{fig/RuntimeExtension-inputpolicy.pdf}
	\caption{Input EDTA $\mathcal{A}_{ped_I}$ obtained from $\mathcal{A}_{ped}$ in Figure \ref{fig:edta-car-rte}}
	\label{fig:inp-edta-car-rte}
	\vspace{-5mm}
\end{figure}

\begin{example}[Input \ac{EDTA} $\calA_{I}$ obtained from $\calA$]
Let us consider the \ac{EDTA} in Figure~\ref{fig:edta-car-rte} defining the property introduced in Example~\ref{eg:edta}.
Figure~\ref{fig:inp-edta-car-rte} presents the input \ac{EDTA} obtained from the \ac{EDTA} in Figure~\ref{fig:edta-car-rte}.
As can be seen, the two upper violation transitions need not be taken as their guard is also satisfied by the self loop on $l_{slow}$ and the transition from $l_{check}$ to $l_{slow}$.
\end{example}
\paragraph{Edit Functions}
Given property $\varphi\subseteq \calD_C^*$, defined as \ac{EDTA} $\calA_{\varphi}=\left(L, {l_0}, X, C,  F,  \Delta \right)$ with semantics $\sem{\calA_\varphi}=\left( Q, q_0, Q_F, \Gamma, \to \right)$,
we introduce $\editI$ (resp. $\editO$), which the enforcer uses for editing input (resp. output) events (whenever necessary), according to input property $\varphi_I$ (resp. input-output property $\varphi$).
Note that in each step the enforcer first processes the input from the environment, and transforms it using $\editI$ based on the input property $\varphi_I$ obtained from the input-output property $\varphi$ that we want to enforce.
Later, the output produced by the program is transformed by the enforcer (when necessary) using $\editO$ based on the input-output property $\varphi$ that we want to enforce.
\squishlist
\item {{\boldmath$\editI\left(\sigma_I\right)$}}:~~Given $I$ (set of input channels), $\editI\left(\sigma_I\right)$ is the set of all possible valuations $\eta_I$ (where $\eta_I$ is a tuple of values of variables in $I$) such that the word obtained by extending $\sigma_I$ with $\eta_I$ can be extended to a sequence that satisfies $\varphi_I$ (i.e., there exists $\sigma'\in \calD_I^*$ such that $\sigma_I\cdot \eta_I \cdot \sigma'$ satisfies $\varphi_I$).
Formally,\\
\[\editI\left(\sigma_I \right) = \{ \eta_I \in \calD_I: \exists \sigma'\in \calD_I^*, \sigma_I \cdot \eta_I \cdot \sigma' \models \varphi_I \}.\]

Consider the automaton
 $\calA_{\varphi_I}=\left(L, {l_0}, X,  C, F,  \Delta_I \right)$ with semantics $\sem{\calA_{\varphi_I}}=\left( Q, q_{0_I}, Q_{F_I}, \Gamma_I, \to_I \right)$.
 Let $q_I\in Q_I$ correspond to a state reachable in $\calA_{\varphi_I}$ (i.e., $q_{0_I} \xrightarrow{\sigma_I} q_I$) upon $\sigma_I$.
 	We define $\editIaut\left(q_I\right)$ as follows:
\[\editIaut\left( q_I \right) = \{ \eta_I \in \calD_I: \exists \sigma'\in \calD_I^*, q_I \xrightarrow{\eta_I \cdot \sigma'}_I q{'{_I}} \wedge q{'{_I}} \in Q_{F_I} \}.\]

\item {\boldmath$\editO\left(\sigma, \eta_I\right)$}:
~~Given an input-output word $\sigma\in \calD_C^*$ and an input event $\eta_I\in \calD_I$, $\editO\left(\sigma, \eta_I\right)$ is the set of valuations of output channels $\eta_O$ in $O$ such that the input-output word obtained by extending $\sigma$ with $(\eta_I,\eta_O)$ can be extended to a sequence that satisfies the property $\varphi$ (i.e., $\exists \sigma' \in \calD_{C}^*$ such that $\sigma\cdot(\eta_I,\eta_O)\cdot\sigma'\models\varphi$).
Formally,
\[\editO\left(\sigma,\eta_I\right) = \{\eta_O \in \calD_O: \exists \sigma'\in \calD_{C}^*, \sigma \cdot \left(\eta_I,\eta_O\right) \cdot \sigma' \models \varphi \}.
\]

Consider the \ac{EDTA} $\calA_{\varphi}\left(L, {l_0}, X, C, F,  \Delta \right)$ defining property $\varphi$ with semantics $\sem{\calA_\varphi}=\left( Q, q_{0}, Q_{F}, \Gamma, \to \right)$, and an input event $\eta_I\in \calD_I$.
If $q\in Q$ corresponds to a state reached in $\calA_{\varphi}$ upon $\sigma$ (i.e., $q_{0} \xrightarrow{\sigma} q$), $\editO\left(\sigma, \eta_I\right)$ can be alternatively defined as follows:
\\
$\editOaut\left(q,\eta_I\right) = \{\eta_O \in \calD_O: \exists \sigma'\in \calD_C^*, q \xrightarrow{\left(\eta_I,\eta_O\right) \cdot \sigma'} q' \wedge q' \in Q_F \}.$

\paragraph{Selecting edits}
For any given violation transition (e.g. $q \xrightarrow{\left(\eta_I,\eta_O\right)} q' \wedge q' \notin Q_F$)
there may be many possible alternate values for $\eta_I$ and $\eta_O$ that would instead result in an accepting transition.
To solve this issue, we define two additional functions, $\selEditI$ and $\selEditO$ which the designer can use to \emph{select} a given edit by the designer from the set of possible accepting edits.

\begin{example}
	\todo{Check, update example!}
Consider the case in Figure~\ref{fig:vdta-car-rte} where the violation transition to $l_v$ occurs if $P \leq 100~\&~B < br\left(P\right)$.
$\editO$ will suggest as valid edits every value $B \geq br\left(P\right)$, as this would invalidate the violation transition guard.
However, this suggestion is infinite in size for real-valued $B$.
As a result, the designer selects edit $B = br\left(P\right)$, which is a valid edit in the solution space.
\end{example}

\item
{\boldmath$\selEditI\left(\sigma_I\right)$:}~~ Given $\sigma_I\in \calD_I^*$ if $\editI\left(\sigma_I\right)$ is non-empty, then $\selEditI\left(\sigma_I\right)$ returns an element (chosen by the designer) from $\editI\left(\sigma_I\right)$, and is undefined if $\editI\left(\sigma_I\right)$ is empty.
Given $q_I\in Q_I$, if $\editIaut\left(q_I\right)$ is non-empty, then $\selEditIaut\left(q_I\right)$ returns an element (chosen by the designer) from $\editIaut\left(q_I\right)$, and is undefined if $\editIaut\left(q_I\right)$ is empty.
\item {\boldmath$\selEditO\left(\sigma,x\right)$:}~~ Given $\sigma\in \calD_{C}^*$, and $x\in \calD_I$,  if $\editO\left(\sigma,x\right)$ is non-empty, then $\selEditO\left(\sigma,x\right)$ returns an element (chosen by the designer) from $\editO\left(\sigma,x\right)$, and is undefined if $\editO\left(\sigma,x\right)$ is empty.
Given $q\in Q$ and $x \in \calD_I$, if $\editOaut\left(q,x\right)$ is non-empty, then $\selEditOaut\left(q,x\right)$ returns an element (chosen by the designer) from $\editOaut\left(q,x\right)$, and is undefined if $\editOaut\left(q,x\right)$ is empty.

\ignore{

\item {\boldmath$\minEditI\left(\sigma_I, x\right)$:}~~ Given $\sigma_I\in \calD_I^*$ and $x\in \calD_I$, if $\editI\left(\sigma_I\right)$ is non-empty, then $\minEditI\left(\sigma_I, x\right)$ returns an event from $\editI\left(\sigma_I\right)$ with minimal distance\footnote{Distance between two events belonging to the same alphabet is the number of bits that differ in both the events.} w.r.t $x$, and is undefined if $\editI(\sigma_I)$ is empty.
Given $q_I\in Q_I$ and $x \in \calD_I$, if $\editIaut\left(q_I\right)$ is non-empty, then $\minEditIaut\left(q_I, x\right)$ returns an event from $\editIaut\left(q_I\right)$ with minimal distance w.r.t $x$, and is undefined $\editIaut\left(q_I\right)$ is empty.
\item {\boldmath$\minEditO\left(\sigma, x, y\right)$:}~~ Given $\sigma\in \calD_{C}^*$, $x\in \calD_I$ and $y\in \calD_O$, if $\editO\left(\sigma,x\right)$ is non-empty, then $\minEditO\left(\sigma, x, y\right)$ returns an event from $\editO\left(\sigma,x\right)$ with minimal distance w.r.t $y$, and is undefined if $\editO\left(\sigma, x\right)$ is empty.
Given $q\in Q$, $x\in \calD_I$ and $y\in \calD_O$, if $\editOaut\left(q,x\right)$ is non-empty, then $\minEditOaut\left(q, x, y\right)$ returns an event from $\editOaut\left(q,x\right)$ with minimal distance w.r.t $y$, and is undefined $\editOaut\left(q,x\right)$ is empty.

}

\squishend
}

\ignore{

\todo{
Old work follows:

An extension wrapper monitors and corrects both input and output of a system according to a given correctness property $\varphi$.
To synthesise a wrapper for a given property $\varphi$ defined as a \ac{EDTA} $\calA_\varphi$, we borrow from the semantics presented in \cite{RuntimeEnforcementOfCPS}.

A wrapper for a property $\varphi$ can only edit an input-output event when necessary, and it cannot block, delay or suppress events.
Let us recall the two functions $\editI$ and $\editO$ that were introduced in Section~\ref{sec:editfunc} that the enforcer for $\varphi$ uses to edit the current input (respectively output) event according to property $\varphi$.

A wrapper for a given property defined as a EDTA $\calA_\varphi$ can be thought as a function $E_{\calA_\varphi}:{\calD}_{C}^* \rightarrow {\calD}_{C}^*$. 
The wrapper aims to keep the property $\calA_\varphi$ satisfied, and so will examine the updated external variables (input and output channels) each tick, and will transform any that are non-accepting.

It can be trivial to derive wrappers for properties which do not behave in a useful manner.
For instance, in the example presented in Figure~\ref{fig:edta-car-rte}, a trivial wrapper would simply keep the speed controller $s$ set to $max\_dec$ at all times.
This would keep the policy satisfied, but it would not result in a useful \ac{AV}.

To prevent this (and other situations), several constraints are provided which define enforcer correctness~\cite{RuntimeEnforcementOfCPS}.

\begin{definition}[Enforcer for $\varphi$]
	\label{def-E-func-constraints}
	Given property $\varphi\subseteq\calD_C^*$, an {\em enforcer} for $\varphi$ is a function $\ef: \calD_C^*\rightarrow \calD_C^*$ satisfying the following constraints:
	
	{\bf Soundness}
	\begin{equation}
	\tag{\bf Snd}\label{eq:snd}
	\forall \sigma \in \calD_C^*, \exists \sigma' \in \calD_C^*:  E_{\varphi}\left(\sigma\right)\cdot \sigma' \models \varphi.
	\end{equation}
	{\bf Monotonicity}
	\begin{equation}
	\tag{\bf Mono}\label{eq:mono}
	\forall \sigma, \sigma' \in \calD_C^*: \sigma\pref \sigma' \Rightarrow \ef\left(\sigma\right) \pref \ef\left(\sigma'\right).
	\end{equation}
	{\bf Instantaneity}
	\begin{equation}
	\tag{\bf Inst}\label{eq:inst}
	\forall \sigma \in \Sigma^*: |\sigma| =  |\ef\left(\sigma\right)|.
	\end{equation}
	{\bf Transparency}
	\begin{equation}
	\tag{\bf Tr}\label{eq:tr}
	\begin{array}{ll}
	\forall \sigma\in \calD_C^*, \forall \eta_I \in \calD_I, \forall \eta_O \in \calD_O, \exists \sigma' \in \calD_C^*:\\
	~~~~~\ef\left(\sigma\right)\cdot\left(\eta_I, \eta_O\right) \cdot \sigma' \models \varphi\\
	~~~~~ \Rightarrow \ef\left(\sigma\cdot\left(\eta_I, \eta_O\right) \right) = \ef\left(\sigma\right)\cdot\left(\eta_I, \eta_O\right) .
	\end{array}
	\end{equation}
	{\bf Causality}
	\begin{equation}
	\tag{\bf Ca}\label{eq:ca}
	\begin{array}{ll}
	\forall \sigma \in \calD_C^*, \forall \eta_I \in \calD_I,\forall \eta_O \in \calD_O,\\
	~~~\exists \eta_I' \in \editI\left(\left(\ef\left(\sigma\right)\right)_I\right), \exists \eta_O' \in \editO\left(\ef\left(\sigma\right), \eta_I'\right):\\
	~~~~~\ef\left(\sigma\cdot\left(\eta_I,\eta_O\right)\right)= \ef\left(\sigma\right)\cdot\left(\eta_I',\eta_O'\right).
	\end{array}
	\end{equation}
\end{definition}

\squishlist
\item A wrapper must be \textit{sound}, meaning that for any word $\sigma$ given as input, the output of the wrapper $E_\varphi\left(\sigma \right)$ should satisfy the property $\varphi$.
\item A wrapper must be \textit{transparent}, meaning that the wrapper must edit the actual input and output channel values only when necessary (i.e., only when they lead to violation of the property).
\item A wrapper is \textit{online}, so it cannot undo what is released as output (\textit{monotonicity}), and it must not delay, insert, or suppress ticks (\textit{instantaneity}).
\item A wrapper must be \textit{causal}, meaning that the enforcer must act as an intermediary such
that it first examines values of the input channels to validate them w.r.t property $\varphi$. If the actual input values will lead to a violation, the wrapper may augment the inputs before
forwarding to the controller. After the controller reacts to these inputs, the wrapper must again validate the outputs from the controller. It should forward the outputs from the controller (without editing) to the environment if they do not lead to a violation, or alter and forward the altered values.  
\squishend
\begin{definition}[Enforceability]
	Let $\varphi\subseteq\calD_C^*$ be a property. We say that $\varphi$ is {\em enforceable} iff a wrapper $\ef$ for $\varphi$ exists according to Definition~\ref{def-E-func-constraints}.
\end{definition}

}
}

\section{RI Manager Synthesis}
\label{sec:synthesis}

\subsection{Problem Definition}
\label{sec:def}

\begin{figure}[!htbp]
	\centering
	\vspace{-3mm}
	\includegraphics[width=0.45\textwidth]{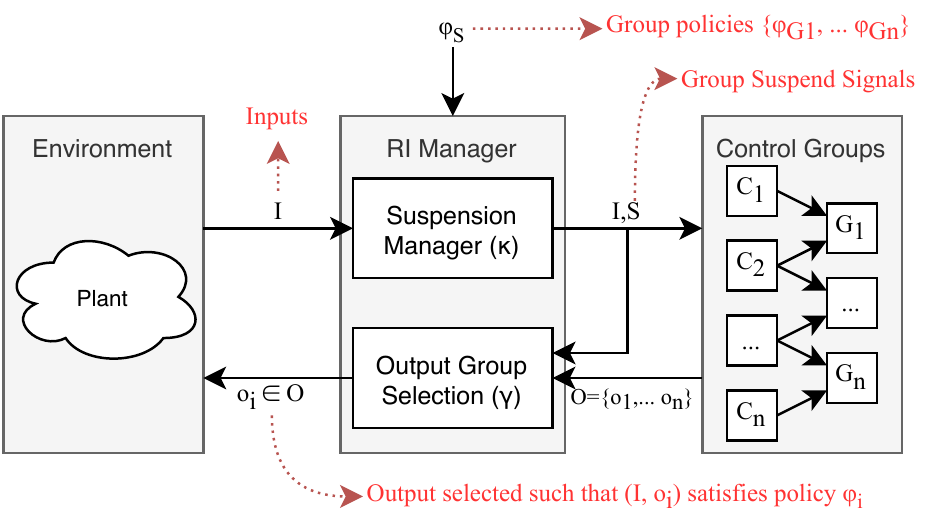}
	\vspace{-3mm}
	\caption{Context of RI Manager}
	\label{fig:RImanager}
\end{figure}

We consider the system (controller) as a \emph{grey-box}, since the RI Manager takes into account the number of controller groups present within the controller, though the internals of each controller is considered to be unknown. 
The context of the RI Manager is illustrated in Figure \ref{fig:RImanager}.
For the synthesis of the RI Manager, the user (designer) provides a set of policies, one policy per controller group defined as VDTA. 
 
Given a set of policies denoted as $\varphi_S = \{\varphi_1,\cdots, \varphi_n \}$ where $n$ is the number of controller groups, $\forall i \in 1\cdots n: \varphi_i \subseteq \calD_C^*$ is the policy corresponding to controller group $i$.

It every reaction (tick), the input-output to the RI Manager is a pair $(x,(y_1, \cdots y_n))$, where $x\in\calD_I$ is the input and $(y_1, \cdots y_n)\in\calD_O \times \calD_O \cdots \times \calD_O$ is an ordered $n$ tuple where $y_i \in \calD_O$ is the output from controller group $i$.
Upon consuming an event $(x,(y_1, \cdots y_n))$ as input, the RI Manager produces an input-output event $(x, y_i) \in \calD_C$ as output, where $y_i \in \{y_1, \cdots y_n\}$ and $y_i \neq \bot$.

In every tick, the RI Manager first processes the input $x$ (i.e., input from the environment), then forwards both this input and an n-tuple of Boolean values (one Boolean value per controller group) indicating to the controller groups which groups should be executed in that tick and which should be suspended.     

After the controller groups have finished, it then receives their n-tuple of outputs $(y_1, \cdots, y_n)$ where $y_i \in \calD_O \cup \{\bot\}$, where $y_i$ corresponds to the output produced by controller group $i$. The symbol $\bot$ is used to denote invalid output (used when a controller group is suspended and will not produce any output in a particular tick).
The RI Manager finally selects one output  
$y_i$ from $(y_1, \cdots, y_n)$, and the final input-output event from the RI manager in a tick is $(x,y_i) \in \calD_C$.

\begin{definition}[RI Manager for $\varphi_S = \{\varphi_1, \cdots, \varphi_n \}$]
	\label{def-N-func-constraints}
	Given a set of policies $\varphi_S = \{\varphi_1, \cdots, \varphi_n \}$, where $\varphi_i \subseteq\calD_C^*$, an {\em RI manager} for $\varphi_S$ is a function $M: (\calD_I \times(\calD_{O1} \times \cdots \times \calD_{On}))^* \rightarrow (\calD_I \times \calD_O)^*$ satisfying the following constraints \footnote{Note that the output domain of each controller group is $\calD_O$ (i.e., $\forall i \in [1, n], \calD_{Oi}$ is $\calD_O$).}:
	
	{\bf Soundness}
	\begin{equation}
	\tag{\bf Snd}\label{eq:snd}
	\begin{array}{ll}
	\forall \sigma \in(\calD_I \times(\calD_{O1} \times \cdots \times \calD_{On}))^+, \exists w \in (\calD_I \times \calD_O)^*,\\
	~~~ \exists i \in [1,n]: |w|=|M(\sigma)|-1 \wedge \pi_1(\sigma_{[1,|\sigma|-1]}) =\pi_1(w) \wedge\\ 
	~~~~~~~~~~~~~~~ \pi_2(M(\sigma)_{[|M(\sigma)|]}) \neq \bot \wedge w\cdot M(\sigma)_{[|M(\sigma)|]} \in \varphi_i.
	\end{array}
	\end{equation} 
	{\bf Monotonicity}
	\begin{equation}
	\tag{\bf Mono}\label{eq:mono}
	\forall \sigma, \sigma' \in (\calD_I \times(\calD_{O1} \times \cdots \times \calD_{On}))^*: \sigma\pref \sigma' \Rightarrow M(\sigma) \pref M(\sigma').
	\end{equation}
	{\bf Instantaneity}
	\begin{equation}
	\tag{\bf Inst}\label{eq:inst}
	\forall \sigma \in (\calD_I \times(\calD_{O1} \times \cdots \times \calD_{On}))^*: |\sigma| =  |M(\sigma)|.
	\end{equation}
	{\bf Causality}
	\begin{equation}
	\tag{\bf Ca}\label{eq:ca}
	\begin{array}{ll}
	\forall \sigma \in (\calD_I \times(\calD_{O1} \times \cdots \times \calD_{On}))^*, \\
	~~~\forall x \in \calD_I,\forall (y_1, \cdots, y_n) \in (\calD_{O1} \times \cdots \times \calD_{On}):\\
	~~~~\exists i \in [1,n]: 
	\sigma_I \cdot x \in \varphi_{i{_I}} \wedge y_i \neq \bot \wedge~IO(\sigma_I, \sigma_O{_i})\cdot (x, y_i) \models \varphi_i.
	\end{array}
	\end{equation} 
\end{definition}

\paragraph{Soundness}
This constraint demands that the output released in each tick should be sound w.r.t one of the policies, and the policy that is satisfied by the final output of the RI manager in each tick may be different (permitting interchanging the controller groups).  

(\ref{eq:snd}): 
For any input word $\sigma$, the output of the RI manager is $M(\sigma)$, and the last event in the output is $M(\sigma)_{[|M(\sigma)|]}$. 

The output in the input-output event $M(\sigma)_{[|M(\sigma)|]}$ should not be $\bot$ (i.e., $\pi_2(M(\sigma)_{[|M(\sigma)|]}) \neq \bot$).

The last event that is released as output ($M(\sigma)_{[|M(\sigma)|]}$) should be sound w.r.t. one of the policies $i \in [1,n]$. 
Let $w\in (\calD_I \times \calD_O)^*$ denote the input-output of controller group $i$ in the previous ticks. 
\begin{itemize} 
\item The length of $w$ should be $|M(\sigma)| -1$.
\item The inputs in $w$ should match with the inputs in $\sigma$ (i.e., $\pi_1(\sigma_{[1,|\sigma|-1]}) =\pi_1(w)$). In the previous ticks, the outputs in $w$ may be different from the outputs in $M (\sigma)$ because the outputs in $M(\sigma)$ may be sound w.r.t policy different from $i$ in the previous ticks. 
\item  The word that is obtained by concatenating $w$ with the last event that is released as output should satisfy policy $\varphi_i$ (i.e.,  $w\cdot M(\sigma)_{[|M(\sigma)|]} \in \varphi_i$). 
\end{itemize}

\ignore{
(\ref{eq:snd}) means that for any input word $\sigma$, where the output of the RI manager $M(\sigma)$, the last event that is released as output by the RI manager $M(\sigma)_{[|M(\sigma)|]}$ should be satisfying at least one of the policies $i \in [1,n]$ w.r.t the previous output of the controller group $i$ (i.e., the outputs produced by it in the previous ticks, for the inputs in $\pi_1(\sigma_{[1,|\sigma|-1]})$) followed by the new event released as output by the RI manager:
\begin{itemize}
	\item the output in the input-output event $M(\sigma)_{[|M(\sigma)|]}$ should not be $\bot$ (i.e., $\pi_2(M(\sigma)_{[|M(\sigma)|]}) \neq \bot$)
	\item there should be an input-output word $w$ which is of length $|M(\sigma)|-1$, where the inputs in $w$ match with the inputs in $M(\sigma)$, and the word that is obtained by concatenating the output last event of the RI manager with $w$ should satisfy policy $\varphi_i$.   
\end{itemize} 
}

\paragraph{Monotonicity}
The monotonicity constraint expresses that  what is already released as output by the RI manager cannot undone.
(\ref{eq:mono}) defines that the output of the RI manager for an extended input word $\sigma'$ of an input word $\sigma$, extends the output produced by the RI manager for $\sigma$.

\paragraph{Instantaneity}
The instantaneity constraint means that the RI manager cannot suppress, delay and insert events.
(\ref{eq:inst}) expresses that for any given input sequence $\sigma$, the output of the RI manager $M(\sigma)$ should contain exactly the same number of events that are in $\sigma$ (i.e., $|\sigma| = |M(\sigma)|$).
This means that, 
in every tick, RI manager receives a new event, and it must react instantaneously and produce an output event immediately.
This requirement is essential for \acp{CPS}, which are reactive in nature.
\paragraph{Causality}
(\ref{eq:ca}) expresses that for every new  event $(x,(y_1, \cdots, y_n))$ the RI manager first processes the input part $x$, to produce Boolean signals which will be sent to the controller along with the input $x$ indicating the controller regarding which controller groups have to be executed. There should be at least one non-suspended controller group policy $i \in [1,n]$, such that the input word obtained by concatenating the new input event $x$ to $\sigma_I$ (previous input) will satisfy the input property corresponding to policy $\varphi_i$. 
Moreover, the output produced by controller group $i$ should not be $\bot$ (controller group should not be suspended), and the input-output word of the controller group $i$ for $\sigma$ (which is $IO(\sigma_I, \sigma_O{_i})$) followed by $(x, y_i)$ (which is the new input-output of controller group $i$) should satisfy the policy $\varphi_i$.

\subsection{Functional Definition}
\label{sec:func:def}
We now provide a definition of a RI Manager for a given set of policies $\varphi_S = \{\varphi_1,\cdots, \varphi_n \}$.

\begin{definition}[RI Manager Function]
	\label{def-func-RI} Given a set of properties $\varphi_S = \{\varphi_1, \cdots, \varphi_n\}$, where $\varphi_i\subseteq\calD_C^*$ defined as VDTA $\calA_{\varphi_i}$, the RI Manager function $M:(\calD_I^* \times(\calD_{O1} \times \cdots \times \calD_{On})^*) \rightarrow (\calD_I \times \calD_O)^*$ is defined as:
	\[
	\begin{array}{lll}
	M(\sigma_I, (\sigma_{O1}, \cdots, \sigma_{on})) \!\!& = \gamma(\kappa(\sigma_I), (\sigma_{O1}, \cdots, \sigma_{on}))
	\end{array}
	\]
	where:
	$\kappa: \calD_I^* \rightarrow (\calD_I \times \bbb_I)^* $ is defined as:
	
	\[
	\begin{array}{rll}
	\kappa(\epsilon_{ \calD_I}) & = (\epsilon_{\calD_I}, \epsilon_{\bbb_I}) \\
	\vspace{0.5em}
	\kappa(\sigma_I \cdot x) & =
	\begin{cases}
	\kappa(\sigma_I) \cdot (x, (b_1, \cdots, b_n)) & 
	\\
	\begin{aligned}
&&	\mbox{s.t.}\ & \forall i \in[1,n]: ((b_i== \true)\Rightarrow\\
&& &(\exists \sigma' \in \calD_I^* : \sigma_I \cdot x \cdot \sigma' \in \varphi_{i_{I}})) 
	\end{aligned}
	\end{cases}
	\end{array}
	\]

	\vspace{1em}
	$\gamma: (\calD_I \times \bbb_I)^* \times (\calD_{O1} \times \cdots \times \calD_{On})^*  \rightarrow (\calD_I \times \calD_O)^*$ is defined as:
	\[
	\begin{array}{rll}
	\gamma((\epsilon_{I}, \epsilon_{\bbb_I}), (\epsilon_{o1}, \cdots, \epsilon_{on})) & = (\epsilon_I, \epsilon_O)\\
	\end{array}
	\vspace{1em}
	\]
	\[	
	\begin{array}{rll}
	\gamma(\sigma_{IB} \cdot (x, (b_1,.., b_n)), \sigma_O \cdot (y_1,..,y_n)) & =
	\begin{aligned}
	\gamma(\sigma_{IB}, \sigma_O) \cdot (x, y_i)
	\end{aligned}
	
	\end{array}
	\]
	
		\vspace{-2em}
 \[~~~~~~~~\mbox{s.t.}\   b_i == \true \wedge y_i \neq \bot \wedge IO(\pi_1(\sigma_{IB}), \pi_i(\sigma_{O}))\cdot(x, y_i) \in \varphi_i.\]

\end{definition}
Let us understand Definition~\ref{def-func-RI} further.
Function $M$ (RI Manager) takes a word over $(\calD_I \times(\calD_{O1} \times \cdots \times \calD_{On}))^*$ and returns a word over $(\calD_I \times \calD_O)^*$ as output.
Here $\sigma_I \in \calD_I^*$, and $\forall i \in \bbn: \sigma_{oi} \in \calD_O \cup \{\bot \}$. 

Function $M$ is defined as a composition of two functions $\kappa$ and $\gamma$, where the function $\kappa$ is to check each policy w.r.t the input received from the environment and to indicate the controller of which controller groups have to be executed. 
Function $\gamma$ takes the output from all the controller groups and selects one among them to be forwarded to the environment as the final output of the RI system.

\paragraph{Function $\kappa$}
Function $\kappa$ takes a word  $\sigma_I\in\calD_I^*$ as input and returns a word in $(\Sigma_I \times \bbb_n)^*$ as output.
Function $\kappa$ is defined inductively. It returns $(\epsilon_{\calD_I}, \epsilon_{\bbb_n})$ when the input $\sigma_I = \epsilon_{\Sigma_I}$.
If $\sigma_I$ is read as input $\gamma(\sigma_I)$ is returned as output, and when another new input $x \in \calD_I$ is observed, a new event $(x, (b_1, \cdots, b_n))$ will be appended to the output of the function $\gamma$ where, $b_i$ corresponding to policy $i$ will be $\true$ if the input word $\sigma_i \cdot x$ can be extended to a word that will satisfy the input property corresponding to $\varphi_i$, and will be $\false$ otherwise. 

When the controller receives the event $(x, (b_1, \cdots, b_n))$, controller group $i$ will be executing only if the Boolean signal $b_i$ will be $\true$, and will be suspended if $b_i$ is $\false$. 
  
\paragraph{Function $\gamma$}
Function $\gamma$ takes an input word belonging to $(\calD_I \times \bbb_n)^*$ and an output word belonging to $(\calD_{O1} \times \cdots \times\calD_{On})^*$ as input, and it returns an input-output word belonging to $(\calD_I \times \calD_O)^*$ which is a sequence of tuples, where each event contains an input and an output.
$\sigma_{IB}$ is used to denote a word belonging to  $(\calD_I \times \bbb_n)^*$.

For a given input word $\sigma_I\in\calD_I^*$, the input word fed as input to $\gamma$ is the output of function $\kappa$ (i.e., $\kappa(\sigma_I)$).

Function $\gamma$ is defined inductively.
When both input and output words are empty, the output of $\gamma$ is $\epsilon$.
If $\sigma_{IB} \in (\calD_I \times \bbb_n)^*$, and $\sigma_O \in (\calD_{O1} \times \cdots \times\calD_{On})^*$ is read as input, its output will be $\gamma(\sigma_{IB}, \sigma_O)$,
and when another new input event $(x, (b_1,.., b_n))$ and output event $(y_1, \cdots, y_n)$ is observed, 
the output of function $\gamma$ will be appended with a new event $(x, y_i)$ where
\begin{itemize}
\item $b_i$ is $\true$ and 
\item $y_i \in  (y_1, \cdots, y_n)$ is a valid output (i.e., output different from $\bot$), and	
\item the input-output word obtained from $\pi_1(\sigma_{IB})$ and $\pi_i(\sigma_{O})$ (i.e., $IO(\pi_1(\sigma_{IB}),\pi_i(\sigma_{O}))$)  followed by the event $(x, y_i)$ will satisfy the policy $\varphi_i$. 
\end{itemize}

\begin{proposition}
Given a set of properties $\varphi_S = \{\varphi_1, \cdots, \varphi_n\}$, if an RI manager as per Definition \ref{def-N-func-constraints} exists for $\varphi_S$, then its RI manager function $M$ as per Definition \ref{def-func-RI} satisfies the Soundness, Monotonicity, Instantaneity and Causality constraints as per Definition \ref{def-N-func-constraints}. 
\end{proposition}

Note that an RI Manager does not modify the outputs of the controller groups. 
So, the existence of an RI Manager for a given set of policies depends on the presence of statically verified controller groups (i.e., each group individually satisfies its policies). 
Given such policies and controller groups, an RI Manager will exist. 
The synthesized RI Manager allows the system to dynamically switch among a set of controller groups based on the situation/observed input.   
\subsection{Algorithm}
\label{sec:algo}
In Section~\ref{sec:func:def} we provided an abstract view of our RI Manager defining it as a function $M$.
For a given set of policies $\varphi_S = \{\varphi_1,\cdots,\varphi_n\}$, this function $M$ is defined as a composition of two functions, $\kappa$ and $\gamma$, where function $\kappa$ examines the input from the environment and decides which controller groups should be suspended, and function $\gamma$ examines outputs from all the controller groups, and checks if they are acceptable w.r.t their corresponding policies, finally emitting one output amongst those that were acceptable as the final output to the environment.

In this section we provide an online algorithmic view of our RI Manager defined in Section~\ref{sec:func:def}, which further illustrates how functions $\kappa$ and $\gamma$ and the overall enforcement function $M$ are implemented.

The algorithm requires a set of policies $\varphi_S = \{\varphi_1,\cdots,\varphi_n\}$, (where each policy $i \in [1,n]$ is formally defined as a VDTA $\calA_{\varphi i}=\left(L, {l_0}, F, X, I, O, \Delta \right)$) with semantics $\sem{\calA_{\varphi i}}=\left( Q, q_0, Q_F, \Gamma, \to \right)$  as input. 
Let us recall that for any given property $\varphi_i$ defined as VDTA $\calA_{\varphi i}$, the 
input automaton corresponding to $\calA_{\varphi i}$ (denoted as $\calA_{I_{\varphi i}}$) is obtained from $\calA$ by projecting on inputs.

\begin{algorithm}[ht]
	\caption{\ac{RI} Manager for $\varphi_S = \{\calA_{\varphi1},\cdots, \calA_{\varphi n}\}$}
	\label{algo:rim}
	{
		\begin{algorithmic}[1]
			\State $t \gets 0$
			
			\ForAll{ $i \in [1,n]$}
			\label{algo:InitBegin}
			\State $\mathcal{A_{\varphi\text{i}}}.q \leftarrow\mathcal{A_{\varphi\text{i}}}.q_{0}$
			\EndFor
			\label{algo:InitEnd}
									
			\While {$\true$}
			\State $\eta_{It} \gets \readInp\left( \right)$
			
			\ForAll{$i \in [1,n]$}
			\label{algoK-begin}
			\State $b_i \leftarrow  (\exists  q' \in Q_{F}, \exists \sigma'\in \calD_I^*  \text{ s.t. } \calA_{I_{\varphi i}}.q \xrightarrow{\eta_{It}\cdot \sigma' }_I \calA_{I_{\varphi i}}.q')$
			\EndFor
			\label{algoK-end}
			
			\State $\mathsf{call\_controllers\left(\eta_{It}, (b_i, \cdots, b_n) \right)}$
			\State $(\eta_{O1t}, \cdots, \eta_{Ont}) \gets \readOut\left( \right)$
			
			\ForAll{$i \in [1,n]$}
			\label{algoEo-begin}
			\State $b'_i \leftarrow  ( b_i \wedge \exists  q' \in Q_{F}  \text{ s.t. } \calA_{{\varphi i}}.q \xrightarrow{(\eta_{It}, \eta_{Oit}) } \calA_{{\varphi i}}.q')$
			\label{a1}
			\EndFor
			\label{a2}
			
			\State $\eta_{Ot} \leftarrow \mathsf{sel-NonDet}((b'_1,\cdots, b'_n),(\eta_{O1t}, \cdots, \eta_{Ont}))$
			\label{algoEo-end}
			\State $\release\left(\left(\eta_{It}, \eta_{Ot}\right)\right)$
			\label{algoEo-release}
			
			\ForAll{ $i \in [1,n]$}
			\label{algo-suBegin}
		
			\State $\mathcal{A_{\varphi\text{i}}}.q \leftarrow\mathcal{A_{\varphi\text{i}}}.q'$ ~~~~ {\footnotesize{where $\calA_{{\varphi i}}.q \xrightarrow{(\eta_{It}, \eta_{Oit}) } \calA_{{\varphi i}}.q' $}}
			
			\EndFor
			\label{algo-suEnd}
			
				\State $t \gets t+1$
			\EndWhile
		\end{algorithmic}
	}
\end{algorithm}

Algorithm~\ref{algo:rim} is an infinite loop, and an iteration of the algorithm is triggered at every time step.
We join this algorithm to the controller groups through a \emph{reactive interface}. The runtime algorithm passes data to and from the controller by calling the function  $\mathsf{call\_controller\left(\eta_{It}, (b_1, \cdots, b_i)\right)}$.
Here, $\eta_{It}$ is the observed input, and $\forall i \in [1,n]$,  $b_i$ is a Boolean which indicates the controller group $i$ should be executed with input $\eta_{It}$ if $b_i$ is $\true$, and should be suspended otherwise.

In Algorithm~\ref{algo:rim}, $t$ keeps track of the time-step (i.e. \emph{tick}), and is initialized at 0.
For every controller group $i \in [1,n]$, $\calA_{{\varphi i}}.q$   
keeps track of the state of its policy $\varphi_I$ (i.e., state of both automata $\calA_{\varphi i}$ and $\calA_{I_{\varphi i}}$).
Note that $q$ contains information about the current location $l$, the current valuations of internal variables $\nu$, and the current valuations of the clocks $\chi$.
Lines \ref{algo:InitBegin} to \ref{algo:InitEnd} deals with initializing the state of each policy $\calA_{{\varphi i}}$. 

Functions $\readInp()$ and $\readOut()$ read the input and output channels respectively, and $\release()$ takes an input-output event and emits it from the RI system.
Function $\kappa$ in Definition~\ref{def-func-RI} (which decides which controller groups have to be executed in a tick w.r.t the observed input) corresponds with lines \ref{algoK-begin} to \ref{algoK-end} in Algorithm~\ref{algo:rim}. 
For each controller group $i \in [1,n]$, a Boolean value $b_i$ is computed, which will be $\true$ if it is possible to reach an accepting state in the input automaton now or in the future (corresponding to $\varphi_i$ from the state reached from its current state upon the observed input event $\eta_{It}$).

Function $\gamma$ in  Definition~\ref{def-func-RI}  corresponds to lines \ref{algoEo-begin} to \ref{algoEo-end} in Algorithm \ref{algo:rim}. The for-loop (lines \ref{a1} to \ref{a2}) computes a Boolean $b'_i$ for each controller group $i$, where $b'_i$ will be $\true$ if the controller group $i$ is not suspended and if it is possible to reach an accepting state $q'$ from the current state $q$ in $\calA_{\varphi i}$ upon the event $(\eta_{It}, \eta_{Oit})$ where $\eta_{It}$ is the observed input and $\eta_{Oit}$ is the output produced by controller group $i$. 

Function $\mathsf{sel-NonDet()}$ takes an n-tuple of Boolean signals and an n-tuple of outputs, and non-deterministically selects one output at some index $i \in [1,n]$ for which the corresponding value in the n-tuple of Boolean signals at index $i$ is $\true$.
		
Finally, the current state of each policy is updated (using the input observed and the output produced by the controller group) before proceeding with the next tick (lines \ref{algo-suBegin}-\ref{algo-suEnd}).

\section{Results}
\label{sec:results}
To evaluate the efficacy of the \ac{RI} framework we analyse the performance of the F1/10 case study introduced in Section~\ref{sec:motivating} with the controller and manager network depicted in Figure~\ref{fig:ri-f110}.
We first measured the crash rate of the original Steering MLP from \cite{ivanovCaseHSCC} in the original setting, i.e. a single car driving around an empty course.
We then extend the environment, adding pedestrians that may randomly
appear on the track at any time, and measure the new crash rate.
Finally we add a second and third car, introduced 15 and 30 time units
after the first car respectively, and performed measurements again.
The averaged values across 1,000 trials for each setting are presented in Table~\ref{tbl:res-original} (where multiple cars are in a setting, the crash rate is the average of all cars across all trials). 

\begin{table}[htbp]
	\centering
	\caption{Performance of the original \ac{MLP} controller
	\label{tbl:res-original}}
	\vspace{-2mm}
	\begin{tabular}{|l|l||l|}
		\hline
		\textbf{No. Cars}           & \textbf{No. Peds} & \textbf{Crash Rate} \\ \hline
		\multirow{3}{*}{1} %
		& 0        & 0 \%       
		\\ \cline{2-3} 
		& 1        & 30.8 \%          
		\\ \cline{2-3} 
		& 2        & 53.0 \%          
		\\ \hline
		\multirow{3}{*}{2} %
		& 0        & 0 \%          
		\\ \cline{2-3} 
		& 1        & 44.4 \%          
		\\ \cline{2-3} 
		& 2        & 70.2 \%          
		\\ \hline
		\multirow{3}{*}{3} %
		& 0        & 0 \%          
		\\ \cline{2-3} 
		& 1        & 58.3 \%          
		\\ \cline{2-3} 
		& 2        & 85.5 \%          
		\\ \hline
	\end{tabular}
\vspace{-5mm}

\end{table}

As expected, the original \ac{MLP} is not suitable for control of the autonomous F1/10 in this new environment, with a very high crash rate with the random pedestrians.
We then combine the original Steering MLP with our own custom controllers (the Swerving Heuristic, the Linear Braking model, and the Follower PID) according to the policies laid out in Section~\ref{sec:rex} and synthesise a \ac{RI} Manager according to the methodology in Section~\ref{sec:synthesis}. For implementation purposes, we compile the control network to C.
We then run these modified vehicles in these same environments, again for 1,000 trials per scenario, and present our results in Table~\ref{tbl:res-ri} .

\begin{table}[htbp]
	\centering
	\vspace{-3mm}
	\caption{Performance of the RI Control Scheme \label{tbl:res-ri}}
	\vspace{-2mm}
	\begin{adjustbox}{width=\columnwidth,center}
	\begin{tabular}{|l|l||l|l|l|l||l|}
		\hline
		\textbf{\begin{tabular}[c]{@{}l@{}}No.\\ Cars\end{tabular}}  & 
		\textbf{\begin{tabular}[c]{@{}l@{}}No.\\ Peds\end{tabular}} & \textbf{\begin{tabular}[c]{@{}l@{}}Normal\\ Mode\end{tabular}} & \textbf{\begin{tabular}[c]{@{}l@{}}Stopping\\ Mode\end{tabular}} & \textbf{\begin{tabular}[c]{@{}l@{}}Cautious\\ Mode\end{tabular}} &
		\textbf{\begin{tabular}[c]{@{}l@{}}Change\\ Rate\end{tabular}} & \textbf{\begin{tabular}[c]{@{}l@{}}Crash\\ Rate\end{tabular}} \\ \hline
		\multirow{3}{*}{1} 
		& 0                 
		& 100 \%
		& 0 \%
		& 0 \%
		& 0 \% 
		& 0 \% 
		\\ \cline{2-7} 
		& 1 %
		& 90 \%
		& 2 \%
		& 8 \%
		& 0.8 \% 
		& 0 \%
		\\ \cline{2-7} 
		& 2 %
		& 83 \%
		& 3 \%
		& 14 \%
		& 1.5 \% 
		& 0 \% 
		\\ \hline
		\multirow{3}{*}{2} 
		& 0
		& 100 \%
		& 0 \%
		& 0 \%
		& 0 \%      
		& 0 \% 
		\\ \cline{2-7} 
		& 1 %
		& 89 \%
		& 2 \%
		& 9 \%
		& 0.9 \%       
		& 0 \%
		\\ \cline{2-7}
		& 2 %
		& 81 \% 
		& 3 \%  
		& 16 \% 
		& 1.5 \% 
		& 0 \% 
		\\ \hline
		\multirow{3}{*}{3} 
		& 0
		& 100 \%       
		& 0 \%
		& 0 \%
		& 0 \% 
		& 0 \%
		\\ \cline{2-7} 
		& 1 %
		& 88 \%
		& 2 \%
		& 10 \%
		& 1.0 \% 
		& 0 \% 
		\\ \cline{2-7} 
		& 2 %
		& 81 \%   
		& 3 \%
		& 16 \%
		& 1.6 \% 
		& 0.6 \%                   
		\\ \hline
	\end{tabular}
	\end{adjustbox}
	\vspace{-2mm}
\end{table}

Firstly, we note that the \ac{RI} controllers are orders of magnitude safer than the original controllers in the chaotic environment.
No accidents were recorded for the 1- or 2- car scenarios, and while
accidents were recorded for the 3-car 2-pedestrian situation, these
were rare, and upon examination were caused by pedestrians colliding
with the cars the majority of the time.
That said, given that some crashes were caused by the cars, this
environment is too `busy' for the controllers as designed in
Figure~\ref{fig:ri-f110}, and further controllers 
with additional policies are required for ensuring safety under all
circumstances (except when pedestrians hit cars intentionally).

Table~\ref{tbl:res-ri} also presents the time the \ac{RI} manager selected each mode.
Given no pedestrians, the cars remained in \textit{normal}, and used the pure control scheme, giving the same performance as the pure results in Table~\ref{tbl:res-original} (i.e. no crashes).
Once we add pedestrians, the cars begin switching modes.
The \emph{normal} mode remains active most of the time, but the \emph{cautious} mode, where cars will slow down according to the ``Distance PID'', is also selected for use up to 16 \% of the time. Finally, the \emph{stopping} mode, which brings the cars to a complete halt, is active up to 3 \% of the time.
These controller mode selection ratios largely track one another from scenario to scenario given the same number of pedestrians.
This is because it is most common for the leading car to detect a pedestrian and change mode/slow down, which then causes any following cars to also change mode and slow down.
It is only in rare cases where the leading car(s) are able to avoid a pedestrian without stopping (e.g. via swerving) leaving follower car(s) to stop without the leader.

Finally, in Table~\ref{tbl:res-ri}, the Change Rate column refers to the percentage of operation cycles that resulted in a mode change from one mode to another. As can be seen, mode changes are relatively uncommon, happening only in up to \textbf{1.6 \%} of cycles, meaning that operation according to the policies laid out in Figure~\ref{fig:vdta} is largely stable.

\subsection{Overhead of \ac{RI}}
To examine the overhead of our approach, we measured the execution in CPU cycles (using C's \texttt{<time.h>}) of the \ac{RI} Manager code compared to the combined execution cycles of the controller components. 
We averaged this across all execution scenarios across all sets of 1,000 trials, and found an average introduced overhead of \textbf{2.2 \%}.

\subsection{Scalability of \ac{RI}}
Given that Algorithm~\ref{algo:rim} has
operations structured over four ``for'' loops which iterate over the set of policies, the algorithm will scale in $\mathcal{O}(n)$ time where $n$ is the number of policies.
To verify this we modify the implementation of the \ac{RI} manager to feature $1\times10^5$, $2\times10^5$, $3\times10^5$, and $4\times10^5$ copies of the three policies, and repeat the 1 car 1 pedestrian experiment.
Compared to $1\times10^5$, $2\times10^5$ executed 2.4 times slower; $3\times10^5$ executed 3.3 times slower; and $4\times10^5$ executed 4.0 times slower, confirming the $\mathcal{O}(n)$ nature of the \ac{RI} system.
\section{Related Work}
\label{sec:related}

With the advent of autonomous systems, which may be classed as
intelligent \ac{CPS}~\cite{de2018safety}, there is a need to consider
alternative design
approaches that are amenable to both model-driven and data-driven techniques~\cite{tripakis2018data,Ding2016Modeling}. The challenges posed by such systems
include the need for ensuring safe operation of the system at all
times, including when the environment is non-static. There has been a
recent resurgence of research efforts based on formal methods~\cite{FormalMethods} (which
are especially suited to model-driven development) being combined with
data-driven approaches such as neural networks~\cite{seshia2016towards}. 
Some approaches have suggested encoding networks which are themselves adaptable to their environment (e.g. using Petri Nets~\cite{Ding2016Modeling}). In the adaptive Petri Net model proposed, in ~\cite{Ding2016Modeling}, adaptation in the model is achieved by considering special transitions with neural network algorithms, and the approach also requires modeling the environment.
Likewise \cite{Chen2016Adaptive} demonstrates using a single neural network for adaptive fault-tolerant control of a model helicopter. 
Other approaches consider static verification, where there has been considerable progress~\cite{ai2}. However, scalability remains a concern when considering extremely complex autonomous systems~\cite{seshia2016towards}. Hence, runtime based formal solutions are
getting recent research traction~\cite{de2018safety}.

\ignore{
\red{As autonomous systems are reactive in nature, we need to consider a
class of techniques called \emph{runtime enforcement} of reactive
systems~\cite{ShieldSynthesis}. }
\srinivas{The above sentence may be replaced with the following:}

\srinivas{
Several formally based runtime enforcement approaches have been proposed \cite{enforceablesecpol,RuntimeNonSafety,FalconeMFR11,timedEnforcement}, which are not suitable for autonomous systems as they are reactive in nature. Our work is  related to the class of runtime enforcement techniques that are suitable for reactive systems \cite{ShieldSynthesis, ShieldRealTime, RuntimeEnforcementOfCPS}. These rely on..
}}

Several formally based runtime enforcement approaches have been
proposed
\cite{enforceablesecpol,RuntimeNonSafety,FalconeMFR11,timedEnforcement}, 
which are not suitable for autonomous systems as they are reactive in
nature. Our work is  related to 
the class of runtime enforcement techniques that are suitable for reactive systems \cite{ShieldSynthesis, ShieldRealTime, RuntimeEnforcementOfCPS}.
These rely on low-overhead wrappers, which
 mediate between the environment and the controller of a reactive
 system, to ensure that the system operates safely at all times by
 ensuring that all user specified policies hold. Whenever the input
 and output of the system lead to non-compliance, the \emph{enforcer}
 alters the input / output streams appropriately. 
Early enforcers for reactive systems~\cite{ShieldSynthesis} were uni-directional. More recently, bi-directional
 enforcement has been developed for both industrial processes~\cite{SmartIOModules} and medical devices~\cite{RuntimeEnforcementOfCPS}. Such
 enforcers have also been used to enforce timed policies over
 autonomous systems~\cite{de2018safety}. However, the approach of runtime enforcement is
 not directly applicable in our setting as run-time adaptation to
 requirement changes is needed~\cite{ivanovCaseHSCC}. 

Our work being based on the re-use of several pre-designed and
pre-verified controllers is similar in spirit to \emph{interface
  theories}~\cite{de2001interface}. These are inspired by assume-guarantee reasoning
for studying component compatibility in a formal setting. In~\cite{de2001interface}
components are state-less and their interfaces describe the
assumptions on the inputs and the associated guarantees on the
output. These have been extended to synchronous components (which are
stateful) in~\cite{tripakis2011theory}. While these provide a formal framework for
component-based system design, the interconnection between components
is static, unlike in our setting, where we can dynamically create different
configurations of the controllers to meet changing requirements.

Our work is likewise similar in spirit to that provided via \textit{control barrier functions}~\cite{cbfs}, which seek to provide a method to synthesize safe controllers directly.
However, control barrier functions differ significantly from RI, as they require open models (as opposed to our own black box compositions) and synthesize a single controller (as opposed to our manager that chooses between available controllers).

\ignore{
\todo{
	\begin{itemize}
	\item Runtime assurance
	\item Runtime enforcement
	\item Runtime verification
	\item Contracts

\end{itemize}
}

\todo{
Our runtime extension framework is inspired by the twin philosophies of \acf{RE}~\cite{theoryRE} and \acf{RA}~\cite{RuntimeAssuranceForComplexCPS}. 
Here, \ac{RE} is a subset of \ac{RA} that focuses on making a system 
compliant to some policy $\varphi$ by modifying and/or re-ordering of events in a
system. Initial techniques were designed for
\emph{transformational systems} where delaying events by buffering
is tolerable. As \ac{CPS} are reactive in nature, new methods have
been developed that enforce a set of timed policies by altering the
input / outputs suitably during the same reactive
cycle~\cite{RuntimeEnforcementOfCPS,SmartIOModules}.

 Esterel also provides an excellent avenue to create complex
 applications using  \emph{synchronous concurrency}. This enables
 several \acp{SNN} to be composed synchronously to
 create neural network ensembles~\cite{Maqsood2004}, which are very useful in
 creating complex AI applications that combine \acp{CNN} with other
 types of \acp{ANN} in a systematic way. 
 
 A given ensemble becomes a synchronous program, which can be combined
 with other synchronous components easily in Esterel as the composition
 of synchronous modules. We use this to combine the controller
 represented as a neural network ensemble with the enforcer. Finally,
 the overall system is composed with the reactive environment using the
 standard approach of reactive interfaces. As Esterel programs are
 automatically compiled to a single reactive function in C, where all
 concurrency is ``compiled away'', the generated code is WCET
 analysable, as shown in~\cite{sann}. However, this part is outside the
 scope of the current work.}
}
\section{Conclusions}
\label{sec:conclusions}

There is recent research momentum in the
direction of safe and adaptive autonomy~\cite{verisig,Tavcar2019Review}, including in designing the experimental F1/10 racing car~\cite{ivanovCaseHSCC}. While model-checking can be used to ensure that the racing car
operates safely in a static environment, for safe autonomy to
be practicable, we need to also consider uncertainties in the
environment. 
In other words, autonomous systems need to be adaptive, able to cope with requirement changes. For this purpose we propose a
method called \acf{RI}. We
envision the requirements to be encoded as a set of
\emph{control modes} and we propose an approach to automatically
synthesise a \ac{RI} Manager, which switches between a set of
controllers to keep the system safe at all times. Our approach is demonstrated to have low overheads and linear scalability.
Our work is
similar in spirit to recent work on the use of runtime enforcers with
drones~\cite{de2018safety} which aim to keep a drone within a safe operating range. However, unlike
their work, we are able to dynamically choose between multiple
controllers.

While our work is towards safe autonomy, it is not devoid of
limitations. First, we have to know all operating modes of
the environment a-priori. What happens when a new mode is encountered,
which was never seen before? Second, can we perform incremental design
i.e. when new policies are introduced, could we use the old \ac{RI}
Manager with a new \ac{RI} Manager? These are open problems still to pursue.

\ignore{
As \acp{CPS} become more powerful and interconnected, their number of possible attack vectors grow.
Existing techniques for securing them are primarily adaptations of conventional approaches to cyber security, such as runtime monitoring of network data and implementing secure access control mechanisms.
However, no digital system can ever be considered fully secure.
Designers must also include mechanisms which can ensure safety even in the presence of sophisticated remote attackers gaining complete control over \ac{CPS} control systems.

In this paper, we proposed the idea of adding a run-time enhancement wrapper to an existing \ac{AI} based \ac{CPS} controller.
By applying this technique to an existing antonymous racing car case study, we extended the racing car's functionality, as well as guaranteed its safety in an uncertain environment.
}

\ignore{
\vspace{-3mm}
\section*{Source Access}
The source codes for the runtime enforcement compiler, as well as all case studies and policies, are available online under the MIT license at
https://github.com/PRETgroup/easy-rte.

\section*{Acknowledgements}
This work has been partially supported by The Ministry of Human Resource Development, Government of India (SPARC~P\#701).

\ifCLASSOPTIONcaptionsoff
  \newpage
\fi

}

\bibliographystyle{IEEEtran}
\bibliography{IEEEabrv,./_lit/hammond}
\ignore{
\begin{IEEEbiography}[{\includegraphics[width=1in,height=1.25in,clip,keepaspectratio]{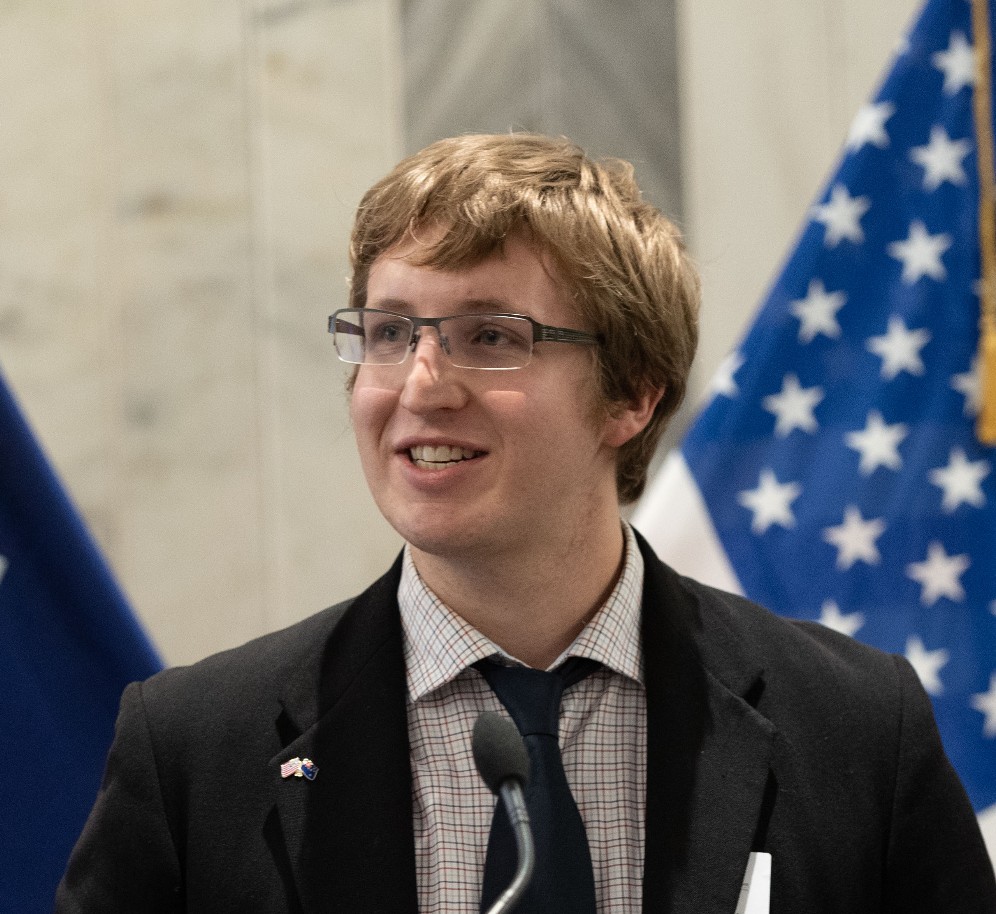}}]\\
	Hammond Pearce received the B.E. (Hons) degree in Computer Systems Engineering from the University of Auckland, Auckland, New Zealand in 2015. 
	He is now pursuing the Ph.D. degree in Computer Systems Engineering at the same institution.
	From 2018 he has been a Professional Teaching Fellow within the Dept. of ECSE at the University of Auckland.
	In 2019 he took part in the NASA International Internship Programme and worked at NASA Ames in California.
	His research interests include IoT, CPS, compilers, industrial automation. 
\end{IEEEbiography}

\begin{IEEEbiography}\\
	Xin Yang ...
\end{IEEEbiography}

\begin{IEEEbiography}[{\includegraphics[width=1in,height=1.25in,clip,keepaspectratio]{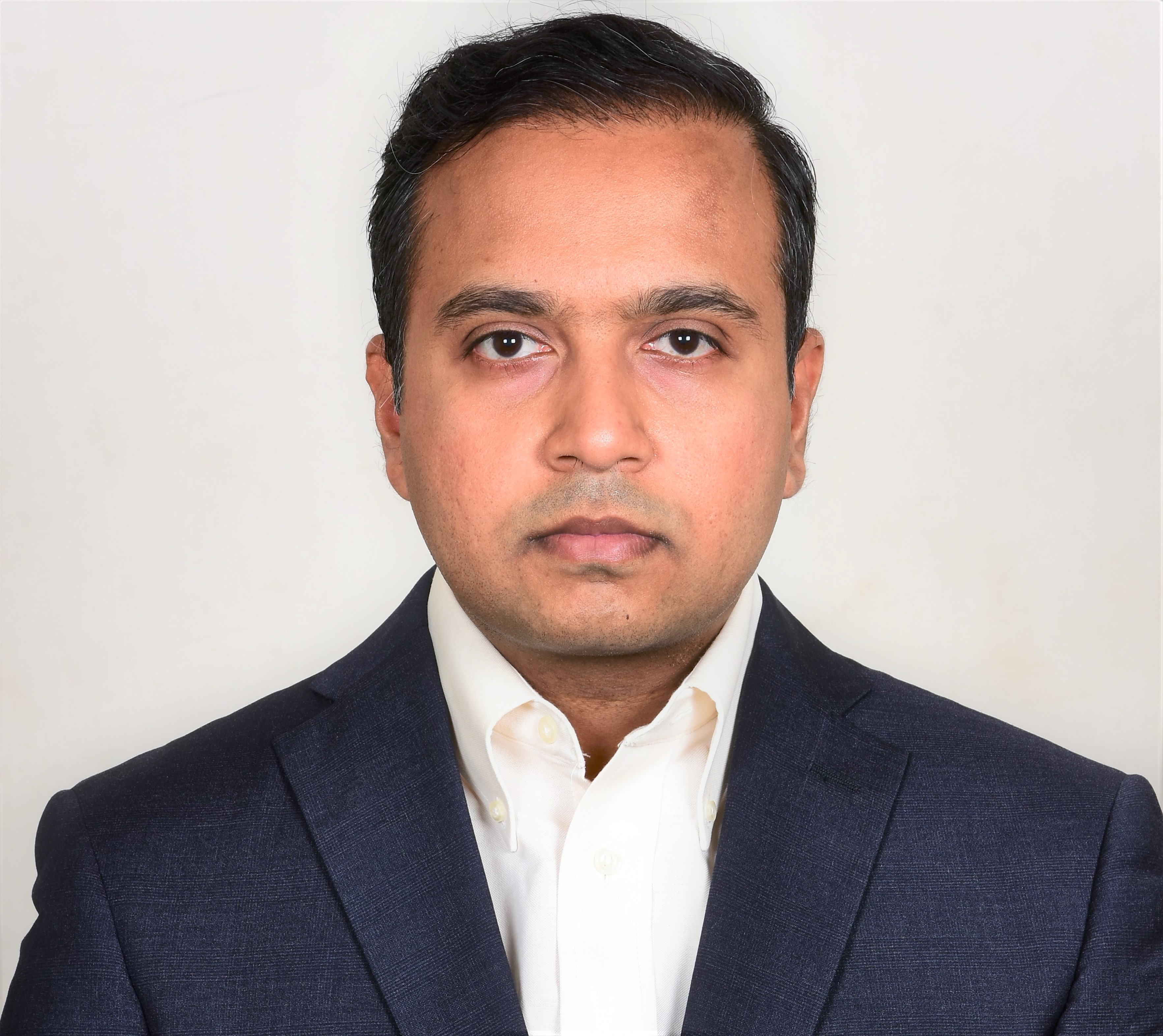}}]\\
	Srinivas Pinisetty received a Ph.D. degree in Computer Science in January 2015 at INRIA, University of Rennes 1, Rennes, France. After completing master's in Computer Science at Eindhoven University Technology (TU/e) in 2009, he continued as a PDEng trainee at TU/e for two years. For his master’s thesis project, he worked at ASML, Veldhoven, Netherlands in 2009, and as a Software Design Engineer trainee at Oc\'e Technologies, Venlo, Netherlands in 2011.
	He is an Assistant Professor in the School of Electrical Sciences at IIT Bhubaneswar. Prior to joining IIT Bhubaneswar he has worked as postdoctoral researcher at the University of Aalto, Finland, and later at the University of Gothenburg | Chalmers, Sweden.
	His research interests include formal methods, and software engineering in general, and runtime verification and enforcement in particular. 
\end{IEEEbiography}

\begin{IEEEbiography}\\
	Partha S. Roop
	Biography text here.
\end{IEEEbiography}
}

\end{document}